\documentclass[10pt,journal,compsoc]{IEEEtran}

\ifCLASSOPTIONcompsoc
  \usepackage[nocompress]{cite}
\else
  \usepackage{cite}
\fi
\ifCLASSINFOpdf
   \usepackage[pdftex]{graphicx}
\else
   \usepackage[dvips]{graphicx}
\fi
\usepackage[cmex10]{amsmath}
\usepackage{amsfonts}
\usepackage[linesnumbered, ruled]{algorithm2e}
\usepackage{array}
\usepackage{mdwmath}
\ifCLASSOPTIONcompsoc
  \usepackage[caption=false,font=footnotesize,labelfont=sf,textfont=sf]{subfig}
\else
  \usepackage[caption=false,font=footnotesize]{subfig}
\fi
\usepackage{url}
\usepackage{multirow}  % for multiline table rows
\usepackage{color}
\usepackage{colortbl}
\usepackage{url}
\usepackage{booktabs}
\usepackage{balance}
\usepackage{epstopdf}
\usepackage{todonotes}
\usepackage{threeparttable}

\DeclareMathOperator*{\argmin}{arg\,min}
\newcommand{\mypara}[1]{{\smallskip \noindent \bf #1}\hspace{0.1in}}
\begin{document}
\title{Sampler Design for Bayesian Personalized Ranking by Leveraging View Data}

\author{Jingtao~Ding,
	   Guanghui~Yu,
	   Xiangnan~He,
        Yong~Li,~\IEEEmembership{Senior Member,~IEEE,}
        and~Depeng~Jin,~\IEEEmembership{Member,~IEEE}% <-this % stops a space
\IEEEcompsocitemizethanks{\IEEEcompsocthanksitem J. Ding, G. Yu, Y. Li and D. Jin are with Beijing National Research Center for Information Science and Technology (BNRist) and with Department of Electronic Engineering, Tsinghua University, Beijing 100084, China. E-mail: \{dingjt15, ygh15\}@mails.tsinghua.edu.cn, \{liyong07, jindp\}@tsinghua.edu.cn.
}
\IEEEcompsocitemizethanks{\IEEEcompsocthanksitem X. He is with School of Computing, National University of Singapore, Singapore. E-mail: xiangnanhe@gmail.com}
}

\IEEEtitleabstractindextext{%
\begin{abstract}
Bayesian Personalized Ranking~(BPR) is a representative pairwise learning method for optimizing recommendation models. It is widely known that the performance of BPR depends largely on the quality of negative sampler. In this paper, we make two contributions with respect to BPR. First, we find that sampling negative items from the whole space is unnecessary and may even degrade the performance. Second, focusing on the purchase feedback of E-commerce, we propose an effective sampler for BPR by leveraging the additional view data. In our proposed sampler, users' viewed interactions are considered as an intermediate feedback between those purchased and unobserved interactions. The pairwise rankings of user preference among these three types of interactions are jointly learned, and a user-oriented weighting strategy is considered during learning process, which is more effective and flexible.
Compared to the vanilla BPR that applies a uniform sampler on all candidates, our view-enhanced sampler enhances BPR with a relative improvement over $37.03\%$ and $16.40\%$ on two real-world datasets. Our study demonstrates the importance of considering users' additional feedback when modeling their preference on different items, which avoids sampling negative items indiscriminately and inefficiently.
\end{abstract}

}

\maketitle

\IEEEdisplaynontitleabstractindextext

\IEEEpeerreviewmaketitle

\section{Introduction}
Due to the prevalence of user implicit feedback in online information systems, recent research on recommendation has shifted from explicit ratings to implicit feedback, such as purchases, clicks, watches and so on~\cite{NCF,iCD}. Different from the recommendation with explicit
ratings~\cite{SVD++,timeSVD++}, negative feedback is naturally scarce when dealing with implicit feedback, also known as one-class problem~\cite{OCF}.
 To learn recommender models from binary implicit feedback, Rendle et al. \cite{BPR} proposed the Bayesian Personalized Ranking~(BPR) method, which assumes that an observed interaction should be predicted with a higher score than its unobserved counterparts (i.e., the missing interactions). 
The optimization of BPR is usually achieved by the stochastic gradient descent (SGD). In each step, it first randomly draws an observed interaction $(u,i)$, and then selects an item $j$ that $u$ has not interacted with before to constitute $(u,i,j)$. Such a process of selecting $j$ is also known as \textit{negative sampling}. 

In the original paper of BPR \cite{BPR}, Rendle \textit{et al.} applied a uniform negative sampler, i.e., sampling $j$ from \textbf{all items} that $u$ has not consumed before with an \textbf{equal probability}. Later on, it was reported that such a uniform negative sampler is highly ineffective and slows down the convergence of BPR~\cite{rendle2014improving,DNS}, especially for datasets that have a large number of items. To this end, \cite{DNS} proposed dynamic negative sampling (DNS) strategies, aiming to maximize the utility of a gradient step by choosing ``difficult'' negative examples --- i.e., the negative examples that lead to a large prediction loss by the current model. 
This process is first randomly selecting $X$ candidates and then drawing a ``difficult'' negative sample with a multinomial distribution based on their prediction scores, where the one with the higher score, i.e., the higher prediction loss, is more likely to be selected. Following this idea of DNS strategy, \cite{rendle2014improving} further proposed a  context-dependent sampler that oversamples informative pairs in each step, and developed an efficient implementation with constant amortized runtime costs.
Despite the significant improvements have been observed, existing DNS strategies sample negative items from the whole item space, which arguably may still suffer from low efficiency when the number of items is large. 

To further mitigate the one-class problem, one intuition is to leverage more side information for learning a more precise preference between two items. In today's implicit recommender systems, besides the primary feedback that can be directly utilized to optimize the conversion
rate, other additional feedback is readily available~\cite{ensemble,empirical}.
Like in E-commerce systems, users' multiple micro-behaviors including view, purchase, wish and put-in-cart are collected~\cite{zhou2018micro}. Similarly,  there are heterogeneous signals related to users' search and watch hitory in online video streaming systems~\cite{covington2016deep}.
Compared to the primary one, the additional feedback always reflects a relative lower level of preference, which could help in learning user preference. For example, in E-commerce systems, user usually views an item before purchasing it. Even though not purchased, a viewed item should still be treated differently when compared with other missing items. 
Also, searching a specific video in online video streaming systems can also be considered as a relatively weak signal of user preference.
As the BPR learns a pairwise ranking relation of user preference between two items, the above additional information can be seamlessly integrated into it by designing an improved BPR sampler.

In this work, we aim to answer the following two research questions: 1) is it necessary to sample negative items from the whole space? and 2) can we design a better sampler for BPR? For the first question about inefficient sampling from whole negative item space, we propose to sample negative items from a reduced space, given that one user normally interacts with a few items. More specifically, a smaller subset is uniformly drawn from the all unobserved items for each user and then fixed as the candidate itemset in the following SGD iterations.
As for the second question, focusing on a specific domain of online-shopping recommender systems, we propose a view-enhanced BPR sampler that considering users' viewed interactions as an intermediate feedback between purchased and unobserved~(i.e., neither purchased nor viewed) interactions. 
We first design a biased sampling process that assumes two-fold semantics in a viewed item, i.e., a negative signal when it was sampled together with another purchased item and a positive signal when with another unobserved item. By tuning the corresponding probability in this biased sampling, the trade-off between these two semantics of user's view signal can be achieved. 
Then, we improve the above scheme by learning the three pairwise ranking relations among a purchased item, a viewed item and an unobserved item together in each training example. In particular, we design a novel objective function with weighted loss to encode the above three relations in the BPR sampler. 
We further assign the weight of these relations based on users' habits in online-shopping activities, which is arguably more effective than the previous methods~\cite{BPR,BPR++,MCBPR} that are limited by the uniformity assumption.

We summarize our key contributions of this work as follows.
\begin{itemize}
\item[1.] We propose to sample negative items from a randomly reduced item space in BPR, and empirically demonstrate that it is unnecessary to sample from all items. When the space is reduced to $1/2^6$ of original size, it achieves a relative improvement of 1.93\% on a popularity-skewed dataset, and only degrades performance within 1.00\% on another less skewed dataset.
\item[2.] We design a view-enhanced user-oriented BPR sampler that can effectively integrate users' viewing data in online-shopping recommender systems, where the viewed interactions are considered as an intermediate feedback between those purchased and unobserved interactions.
\item[3.] We conduct extensive experiments on two real-world datasets, showing that our view-enhanced sampler enhances BPR with a relative improvement of $37.03\%$ and $16.40\%$.
\end{itemize}

The rest of this paper is organised as follows. We review related literature in Section~\ref{sec:related}. Then, we introduce
the dataset and experimental settings in Section~\ref{sec:setup}. The two research questions are investigated in Section~\ref{sec:exp1} and Section~\ref{sec:exp2}, respectively. Finally, we conclude this work and discuss future work in Section~\ref{sec:future}.

\section{Related Work}\label{sec:related}
As implicit feedback data is more common and valuable in modern recommender systems, we first review some related works on modeling user preference from implicit data. Then, we discuss two types of methods that are proposed to improve implicit recommender systems with multiple feedback.

\mypara{Implicit Feedback Systems.}
Handling missing data is notoriously difficult for recommendation with implicit feedback. To solve this problem, two strategies are proposed: whole-data based strategy and sample-based strategy.
Whole-data based strategy treats all missing data as negative feedback~\cite{ALS,eALS,NCF}, while sample-based learning strategy overcomes this problem by sampling negative instances from missing data~\cite{OCF,BPR}.
Both methods have pros and cons: whole-based methods model the full data with a potentially higher coverage, but inefficiency can be an issue; sample-based methods are more efficient by reducing negative examples in training, but risk decreasing the model's performance. As a well-known sample-based method, BPR has been used in many implicit feedback systems. Therefore, in this paper, we focus on developing an improved sampler for BPR. Different from previous works, we demonstrate that 1) it is unnecessary to sample negative items from the whole space, and 2) recommendation performance can be significantly improved after integrating users' additional view data.

\mypara{Collective Matrix Factorization~(CMF).}
CMF is a multiple relational learning method that improves predictive accuracy by sharing information between different feedback~\cite{CMF,GLFM,GroupSparsity}. Originating from the explicit rating problems, it has been extended into implicit case as well~\cite{MRBPR,BF,UTop,MPF,MFPR}.
For example, by applying CMF technique to Bayesian Personalized Ranking, Multi-Relational Factorization with BPR~(MR-BPR) performs better on social network data~\cite{MRBPR}.
A recently proposed method~\cite{MFPR}, namely Multiple Feedback Personalized Ranking~(MFPR), borrows the idea of SVD++~\cite{SVD++} to integrate additional feedback and later optimizes a pairwise ranking loss, which is similar to BPR. 
However, as the CMF-based model generates different user-item relations, i.e., latent factors, for each type of feedback, it is hard to differentiate their preference levels. In contrast, our view-enhanced BPR sampler learns the same user-item relation to indicate relative preference order among purchase and view data, which is more effective.

\mypara{BPR-based Models.}
The second category of methods integrate multiple types of feedback in the sampler of BPR~\cite{BPR++,ABPR,MCBPR}.
The time-based and interaction-count based variants of samplers are designed to provide more signals~\cite{BPR++}. 
From the perspective of transferring knowledge from additional feedback, \cite{ABPR} proposes an adaptive BPR that integrates these feedback to learn better confidence of users' preference on items.
Recently, Multi-channel BPR~(MC-BPR) applies the strategy of assigning different preference levels to multiple types of feedback when sampling training item pairs in BPR~\cite{MCBPR}, which is similar to our proposed view-enhanced scheme based on a biased sampling process. However, by simultaneously modeling pairwise ranking relations among user's purchased, viewed and unobserved items in each training example, our proposed scheme achieves better performance. Moreover, with a user-oriented weighting scheme, the performance can be further improved.

\section{Datasets and Observations}
\label{sec:setup}
\subsection{Datasets}\label{sec:dataset}
We perform experiments on two real-world datasets. 

\textbf{Beibei\footnote{\url{http://www.beibei.com/}}}: Beibei is the largest E-commerce platform for maternal and infant products in China. We sample a subset of user interactions that contain views and purchases from Beibei within the time period from 2017/05/25 to 2017/06/28. 

\textbf{Tmall\footnote{\url{https://www.tmall.com/}}}: Tmall is the largest business-to-consumer E-commerce platform in China. To allow our results to be reproducible, we use a public benchmark released by the ICJAI-2015\footnote{The dataset is downloaded from \url{https://tianchi.aliyun.com/datalab/dataSet.htm?id=5}}. The time period is from 2014/06/01 to 2014/11/11. Note that 11$^{th}$ Nov. of each year is the Tmall Global Shopping Festival\footnote{\url{http://www.alizila.com/look-back-2014-global-shopping-festival/}}, and thus users tend to select many items before and wait for the deals on this day. Therefore, besides the original dataset \textbf{Tmall-all}, we also generate a smaller dataset, called \textbf{Tmall-selected}, to filter out the possible effect brought by this shopping festival, where only those interactions before 2014/10/01 are included.

We take three steps for data preprocessing. We first merge the repetitive purchases of the same user and item into one purchase with the earliest timestamp, as we aim to recommend novel items.
Next we filter out users' views on their purchased items to avoid information leaking. Finally, we filter out users and items with less than 12 and 16 purchases, respectively, to overcome the high sparsity of the raw datasets. Table~\ref{tab_data} summarizes the statistics of our experiment datasets. With both primary~(purchase) and additional~(view) feedback collected, these datasets are sufficient for our research on leverage additional view data in BPR sampler.
\begin{table}[h]
	\renewcommand{\arraystretch}{1.2}
	\setlength\tabcolsep{1.5pt}
	\centering
	\caption{Statistics of the evaluation datasets.}
	\label{tab_data}
	\begin{tabular}{|c|c|c|c|c|c|}
		\hline
		\textbf{Dataset} & \textbf{Purchase\#} & \textbf{View\#} & \textbf{User\#} & \textbf{Item\#} & \textbf{Sparsity} \\ \hline
		Beibei           & 2,654,467           & 23,668,454      & 158,907         & 119,012         & 99.99\%/99.87\%   \\ \hline
		Tmall-all            & 352,768           & 1,585,225       & 28,059          & 32,339          & 99.96\%/99.83\%   \\ \hline
		Tmall-selected            & 160,840             & 531,640       & 12,921          & 22,570         & 99.94\%/99.82\%   \\ \hline
	\end{tabular}
\end{table} 

\subsection{Observations}
\begin{figure}[h]
\centering\hspace{-1em}
\subfloat[purchased interactions]{\includegraphics[width=.255\textwidth]{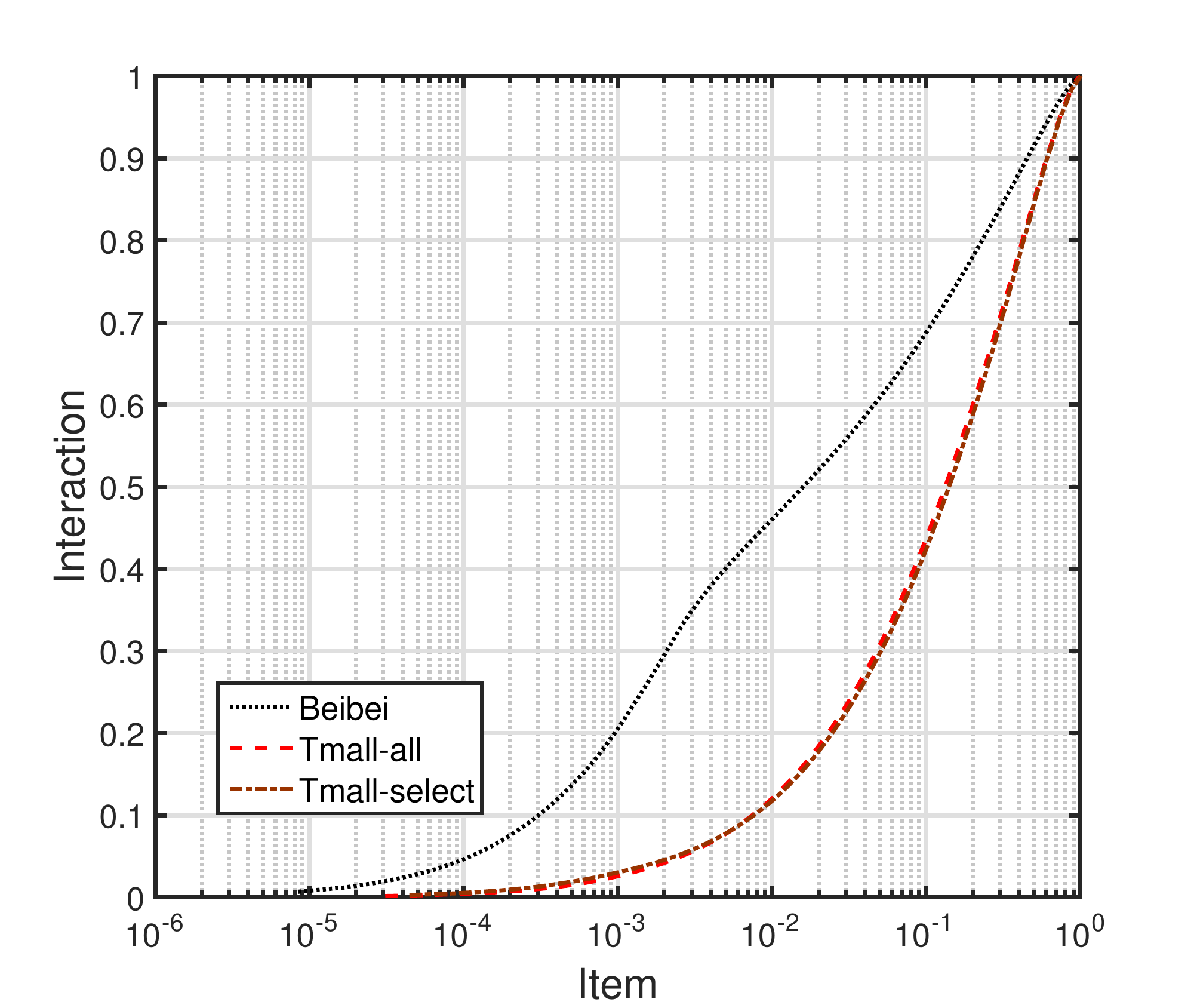}}
\hfill\hspace{-2em}
\subfloat[viewed interactions]{\includegraphics[width=.255\textwidth]{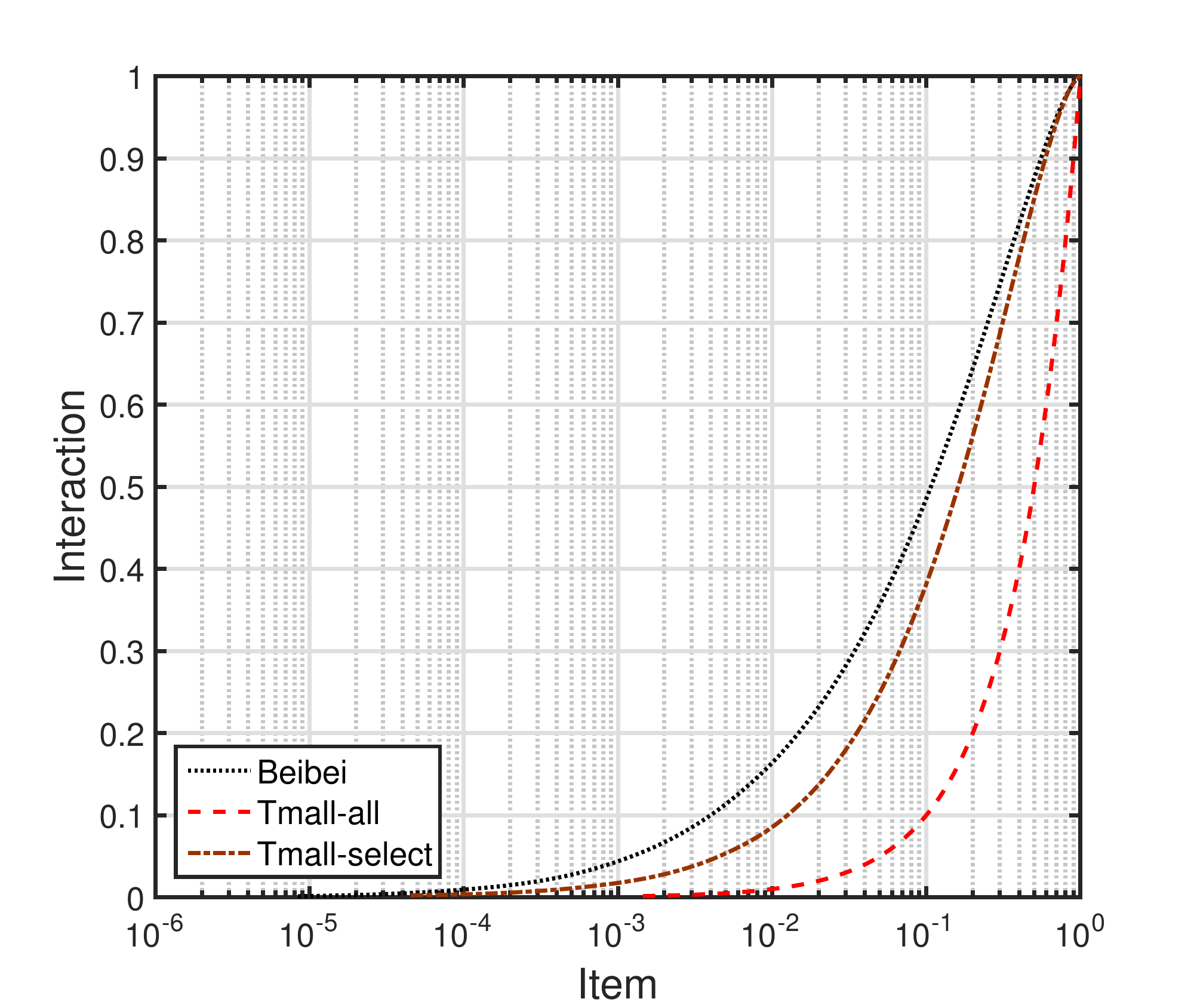}}
\caption{Popularity skewness of the Beibei and Tmall datasets.}
\label{fig_skewness}
\end{figure}

The popularity skewness exists in many recommender systems and impacts the performance.
Therefore, we investigate the popularity skewness in our data, in terms of item purchases and views, and show the result in Fig.~\ref{fig_skewness}(a) and (b), respectively.
The y-axis represents the ratio of interactions for a given ratio of items on the x-axis, sorted by decreasing popularity. 
For item purchases, Beibei is the most popularity skewed dataset, where the top-1\% of the items accounts for 50\% of the purchased interactions, much larger than 10\% in Tmall dataset.
Such difference in skewness no longer exists in item views, where the top-1\% of the items accounts for 16\% and 9\% of the viewed interactions in Beibei and Tmall-selected, respectively.
As for the difference between Tmall-all and Tmall-selected, the popularity skewness of purchase interactions is almost the same, as shown in Fig.~\ref{fig_skewness}(a), while for view interactions Tmall-selected is much more skewed than Tmall-all, about 40\% vs. 10\% in terms of top-10\% of the items.
In summary, users in Beibei are more likely to purchase those popular items, which may affect the performance of personalized recommendation algorithms. On the contrary, users in Tmall-all do not tend to view those popular items, meaning that there may exist a strong personal preference in users' views.

\subsection{BPR}
The objective function for BPR can be formulated as
\begin{equation}
	\argmin_{\Theta} \sum_{(u,i,j)\in \mathcal{D}} -\ln \sigma(\hat{y}_{ui}(\Theta) - \hat{y}_{uj}(\Theta)),
\label{eq1}
\end{equation}
where $\hat{y}(\Theta)$ is the predictive model, and we use the standard matrix factorization~\cite{BPR} as the predictive model.
$\Theta$ denotes the model parameters, $\sigma(x)=\frac{1}{1+\exp(-x)}$ is the sigmoid function to convert the margin to a probability, and $\mathcal{D}$ denotes the set of pairwise training examples:
$\{(u,i,j)| i\in\mathcal{R}_u^+ \wedge j\notin \mathcal{R}_u^+ \}$,
where $\mathcal{R}_u^+$ denotes the set of items that $u$ has interacted with before. Note that we have omitted the $L_2$ regularization terms for clarity. The optimization of BPR is usually achieved by the stochastic gradient descent (SGD). 

\subsection{Evaluation Methodology}
We adopt the \emph{leave-one-out} protocol~\cite{BPR,NCF}, where the latest purchase interaction of each user is held out for testing. 
For hyperparameter tuning, we randomly sample one purchase interaction for each user as the validation set.
The training process is stopped once we observe increasing in the validation loss.

For evaluation measures, we employ \emph{Hit Ratio}~(HR) and  \emph{Normalized Discounted Cumulative Gain}~(NDCG). Mathematically, $HR@k$ for each user $u$ is defined as:
\begin{equation}
	HR_u@k =  \left\{\ \begin{array}{lr}  
             1\ ,\ \text{hit in top-k recommendation}\\  
             0\ ,\ \text{else}.\\  
             \end{array}
	\right.
\end{equation}
$NDCG@k$ for each user $u$ is defined as:
\begin{equation}
	NDCG_u@k =  \sum_{p=1}^k \frac{2^{R(u,p)}-1}{\log(p+1)},
\end{equation}
where $R(u, p)$ is the rating assigned by $u$ to the item at the $p^{th}$ position on the ranked list produced for $u$. Here $R(u, p)$ equals 1 if hit and 0 otherwise. Compared to HR, NDCG is very sensitive to the ratings of the highest ranked items.
We truncate the ranked list of non-purchased items at the position of 100, i.e., k=100, and report the average score of all users.
Since the findings are consistent across the number of latent factors $K$, we report the results of $K = 32$ only.

\section{Unnecessary to sample from all items}
\label{sec:exp1}

\begin{algorithm}[b!]
\caption{Proposed scheme of reducing negative sampling space in BPR.}\label{alg0}
\SetKwInOut{Input}{Input}\SetKwInOut{Output}{Output}
\Input {number of users $M$ and items $N$, user-item interaction data $\mathcal{R^{+}}$, reduce ratio $\gamma$}
\Output {$\Theta$}
\For{$u\gets 1$ to $M$} {
//Generate the negative sampling space for each user \\
$\mathcal{R}^{-}_u \gets random\_select(u,N,\gamma)$
}
\While{not reaching convergence}{
// Random sampling\\
$u \gets$ draw a random user from $\mathcal{U}$\\
$i \gets$ draw a random purchased item from $\mathcal{R}^{+}_u$\\
$j \gets$ draw a random negative item from $\mathcal{R}^{-}_u$\\
Compute gradients of $\Theta$ according to BPR\\
Update the above parameters
}
\end{algorithm} 

\subsection{Methodology}
Generally the vanilla BPR samples negative items indiscriminately from the whole set of those unobserved instances. As the negative sampling space of BPR is fairly large for each user in implicit recommender systems, it may not only cause inefficiency issue but also degrade the performance. To overcome this, we design the following scheme of reducing negative sampling space to evaluate whether it is necessary to sample from all items.

As detailed in Algorithm~\ref{alg0}~(Lines:1-4), we randomly assign each user a much smaller but different set of samples $\mathcal{R}_u^{-}$. More specifically, for each user, the negative items are only sampled from a fraction of items, i.e., a randomly reduced item space, given the size ratio $\gamma$ of this reduced space to the original space. Note that $\gamma$  is invariant among users, while the specific items are different. An intuitive implementation of $random\_select(u,N,\gamma)$ function is to uniformly draw $\gamma\times N$ unobserved instances. Also, more complexed scheme can consider combining both reducing sampling space and applying dynamic negative sampling strategy~\cite{DNS}, as well as adaptively generating sampling space for different users.
Then, with each user's $\mathcal{R}^{-}_{u}$ fixed, BPR sampler randomly samples training triples $(u,i,j)$ and updates model parameters in each iteration.
Since the negative items can only be sampled from the reduced item set, this scheme reduces the number of possible training item pairs $\{(i,j)\}$ for $u$, and thus can largely improve efficiency in terms of learning model parameters.

We vary the size ratio $\gamma$ and summarize the performance on two datasets, Beibei and Tmall-all, in Table \ref{tab:all}. 
In order to factor out random effects, for each size, we repeat the experiment five times and report the average score, as well as standard variance. The first row indicates the performance of the original BPR that samples negative items from the whole space. 

\begin{table}[b]
\renewcommand{\arraystretch}{1.2}
\setlength\tabcolsep{2.0pt}
  \centering
  \caption{Performance of BPR with different settings on the fraction of the reduced sampling space. ``Num.'' means the size of sampling space for each user, i.e., Ratio 
  $\times$ Item\#.}
\label{tab:all}
\subfloat[Beibei]{
    \begin{tabular}{|c|c|c|c|c|c|c|c|}
    \hline
    \textbf{Ratio} & \textbf{Num.} & \multicolumn{2}{c|}{\textbf{HR}} & \textbf{$\Delta$HR} & \multicolumn{2}{c|}{\textbf{NDCG}} & \textbf{$\Delta$NDCG} \\
    \hline
    $2^0$     & 119012 & \multicolumn{2}{c|}{0.1094} & 0     & \multicolumn{2}{c|}{0.0251} & 0 \\
    \hline
    $2^{-5}$    & 3719  & \multicolumn{2}{c|}{$0.1116\pm0.0013$} & +2.36\% & \multicolumn{2}{c|}{$0.0258\pm0.0003$} & +2.71\% \\
    \hline
    $2^{-6}$    & 1859  & \multicolumn{2}{c|}{$0.1112\pm0.0020$} & +2.03\% & \multicolumn{2}{c|}{$0.0256\pm0.0004$} & +1.83\% \\
    \hline
    $2^{-7}$    & 930   & \multicolumn{2}{c|}{$0.1103\pm0.0016$} & +1.22\% & \multicolumn{2}{c|}{$0.0255\pm0.0003$} & +1.67\% \\
    \hline
    $2^{-8}$    & 465   & \multicolumn{2}{c|}{$0.1106\pm0.0022$} & +1.50\% & \multicolumn{2}{c|}{$0.0257\pm0.0007$} & +2.31\% \\
    \hline
    $2^{-9}$    & 232   & \multicolumn{2}{c|}{$0.1104\pm0.0015$} & +1.26\% & \multicolumn{2}{c|}{$0.0255\pm0.0003$} & +1.67\% \\
    \hline
    $2^{-10}$   & 116   & \multicolumn{2}{c|}{$0.1105\pm0.0014$} & +1.39\% & \multicolumn{2}{c|}{$0.0257\pm0.0005$} & +2.47\% \\
    \hline
    \end{tabular}%
  \label{tab:beibei}%
  }
  \vfill
\subfloat[Tmall-all]{
    \begin{tabular}{|c|c|c|c|c|c|c|c|}
    \hline
    \textbf{Ratio} & \textbf{Num.} & \multicolumn{2}{c|}{\textbf{HR}} & \textbf{$\Delta$HR} & \multicolumn{2}{c|}{\textbf{NDCG}} & \textbf{$\Delta$NDCG} \\
    \hline
    $2^0$     & 32339 & \multicolumn{2}{c|}{0.0301} & 0     & \multicolumn{2}{c|}{0.0076} & 0 \\
    \hline
    $2^{-2}$    & 8085  & \multicolumn{2}{c|}{$0.0299\pm0.0003$} & -0.66\% & \multicolumn{2}{c|}{$0.0075\pm0.0001$} & -1.05\% \\
    \hline
    $2^{-3}$    & 4042   & \multicolumn{2}{c|}{$0.0300\pm0.0005$} & -0.27\% & \multicolumn{2}{c|}{$0.0076\pm0.0002$} & +0.26\% \\
    \hline
    $2^{-4}$    & 2021   & \multicolumn{2}{c|}{$0.0300\pm0.0002$} & -0.33\% & \multicolumn{2}{c|}{$0.0076\pm0.0001$} & -0.53\% \\
    \hline
    $2^{-5}$    & 1010   & \multicolumn{2}{c|}{$0.0297\pm0.0007$} & -1.33\% & \multicolumn{2}{c|}{$0.0075\pm0.0002$} & -0.79\% \\
    \hline
    $2^{-6}$    & 505    & \multicolumn{2}{c|}{$0.0299\pm0.0004$} & -0.60\% & \multicolumn{2}{c|}{$0.0075\pm0.0001$} & -1.05\% \\
    \hline
    $2^{-7}$    & 253    & \multicolumn{2}{c|}{$0.0295\pm0.0005$} & -2.06\% & \multicolumn{2}{c|}{$0.0074\pm0.0001$} & -2.37\% \\
    \hline
    \end{tabular}%
    
  \label{tab:tmall}%
  }
\end{table}%

\subsection{Results}
Surprisingly on the Beibei dataset, the performance is not decreased but increased after reducing the sampling space. When varying $\gamma$ from $1/2^{5}$ to $1/2^{10}$, we all observe the performance improvement of over $1.20\%$ in terms of both HR and NDCG. 
Even with a rather small $\gamma$ as $1/2^{10}$, where the sampling space of each user only contains $116$ candidates, we still obtain a relative improvement of $1.39\%$~(HR) and $2.47\%$~(NDCG) over the original BPR. 
This finding is novel and encouraging, meaning that sampling from the whole item space is not only unnecessary for BPR, but may even hurt the performance. 

On the Tmall-all dataset, as the original item space is not that large (which is one magnitude smaller), we do not observe improvements by reducing the sampling space. But still, we can see that with a much smaller sampling space, the performance remains the same level as the original BPR. When the size of sampling space is larger than $505$, i.e., $\gamma>1/2^6$, the performance is only degraded within $1.33\%$ in terms of both HR and NDCG. Only when sampling space becomes one magnitude smaller, a significant decrease over $2.00\%$ is observed.
This provides further evidence on the inefficiency of the uniform sampler for BPR. 

Rendle \textit{et al.}~\cite{rendle2014improving} have shown that oversampling popular items as negative feedback underperforms the basic uniform sampler, due to the under-training of those less popular items. Motivated by this, we investigate the different observations on Beibei and Tmall datasets from this aspect. As we have shown in Fig.~\ref{fig_skewness}(a), Beibei is a popularity skewed dataset, where top-1\% of the items accounts for 50\% of the purchased interactions. By fixing a reduced sampling space for each user, a less popular item in this space can receive more gradient steps. Since there are more unpopular items in Beibei dataset, the scheme of reducing sampling space can benefit the SGD learning process and thus perform better.

To summarize, we have demonstrated that the uniform sampler  is unnecessary for BPR and may even degrade the performance in popularity-skewed datasets. When the space is reduced to $1/2^6$ of original size, it achieves a relative improvement of 1.93\% on the popularity-skewed Beibei dataset, and only degrades performance within 1.00\% on another less skewed Tmall dataset. Considering its inefficiency and poor robustness against popularity skewness, we focus on designing a better sampler for BPR in the following sections.

\section{View-Enhanced Sampler}
\label{sec:exp2}
One inherent issue of recommender systems is the natural scarcity of observed data. To overcome this, BPR samples unobserved items as negative feedback. However, since a user can only interact with a limited number of items, sampling process can be inefficient and may even degrade the performance, as we have demonstrated above.
In E-commerce recommender systems, besides the purchase feedback that is directly related to optimizing the conversion rate, the view logs of users are usually much easier to collect and thus can be leveraged to learn user preference. In this section, we design a view-enhanced sampler for BPR. For
readability, we summarize the major notations throughout the
paper in Table~\ref{tab_note}.

\begin{table}[h]
\small
\renewcommand{\arraystretch}{1.5}
\setlength\tabcolsep{1.2pt}
\centering
\caption{List of commonly used notations.}
\label{tab_note}
\begin{tabular}{|c|l|}
\hline
\textbf{Notation}                            & \textbf{Description}                                                                                                                               \\ \hline
$M,N,K$                                      & The numbers of users, items, and factors.                                                                                                          \\
$\textbf{P},\{\textbf{p}_u\}$                & The latent factor matrix and vector for users.                                                                                                     \\
$\textbf{Q},\{\textbf{q}_i\}$                & The latent factor matrix and vector for items.                                                                                                     \\
$\mathcal{S},\mathcal{S}_u$    & \begin{tabular}[c]{@{}l@{}}The sets of all purchased $(u,i)$ pairs,\\ items purchased by $u$. \end{tabular}        \\
$\mathcal{V},\mathcal{V}_u$    & Similar notations for viewed interactions.                                                                                                           \\
$\mathcal{R},\mathcal{R}_u$    & Similar notations for unobserved interactions.   \\
$\hat{r}_{ui},\hat{r}_{uv},\hat{r}_{uj}$     & \begin{tabular}[c]{@{}l@{}}Predictions of user $u$ over purchased items $i$,\\viewed items $v$ and non-viewed items $j$.\end{tabular}              \\
$\{\omega_{1},\omega_{2},\omega_{3}\}$                                & Probability of sampling training item pairs.                                                                                                       \\
$\alpha$                                        & \begin{tabular}[c]{@{}l@{}}Weight of training pairs made up of\\ a purchased item and a viewed item.\end{tabular}                                                                                                                                                                                                    \\
$\alpha_u$                                        & \begin{tabular}[c]{@{}l@{}}User-oriented weight of training pairs made up of \\a purchased item and a viewed item.\end{tabular}                                                                                                   \\
$\beta$                                        & Significance level of view-purchase ratio in $\alpha_u$.                                                                                                     \\
$\lambda$                                    & Regularization parameter.                                                                                                                          \\ \hline
\end{tabular}
\end{table}

\subsection{Integrating View Signal}
Intuitively, viewed interactions can be treated as an intermediate feedback between the purchased and missing interactions. Therefore, for user $u$'s viewed~(but not purchased) item
$v$, it should have an intermediate value of prediction $\hat{r}_{uv}$ between
those of non-viewed item $j$ (i.e., missing entry) and purchased item $i$, i.e., $\hat{r}_{uj}$ and $\hat{r}_{ui}$.  Based on this, we propose two variant of BPR sampler that can leverage view data. One is to leverage these viewed items in a biased sampling process, the other is to consider this relationship in a newly proposed objective function.
\subsubsection{Biased Sampling}                                                                                                                                                                                                                                                                                                                                                                                                           
\begin{figure}[h]
\centering
\subfloat{\includegraphics[width=.48\textwidth]{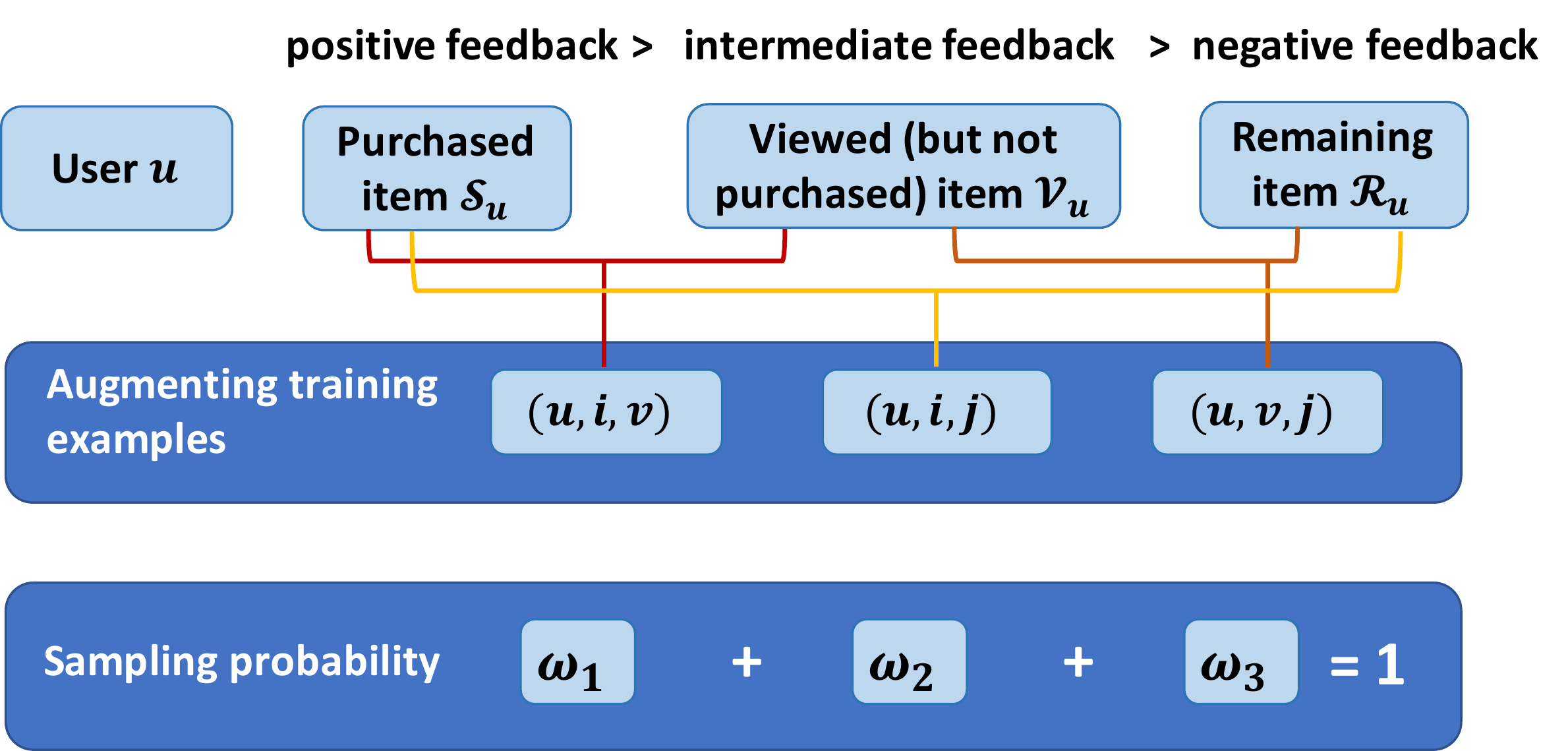}}
\caption{Biased sampling process considering users' viewed items.}
\label{fig_sampling}
\end{figure}

First of all, we can integrate the view signal by augmenting the training data. 
In BPR, a training example $(u,i,j)\in\mathcal{D}$ assumes that $u$ prefers $i$ over $j$. Then, the model parameters, i.e., user vector $\textbf{p}_u$ and item vector $\textbf{q}_i$, are updated towards the objective of $\hat{r}_{ui} > \hat{r}_{uj}$. 
Through a biased sampling process, we are able to encode the intermediate preference information of users' viewed interactions in the model.
In our proposed view-enhanced sampler, as illustrated in Fig.~\ref{fig_sampling}, we split the item space into three sets for each user $u$, namely $\mathcal{S}_u$, $\mathcal{V}_u$, and $\mathcal{R}_u$, which indicate the purchased items, viewed (but not purchased) items, and remaining non-viewed items, respectively. Then, we sample an item pair from three candidate sets,  $\{(i,v)|i\in\mathcal{S}_u,v\in\mathcal{V}_u\}$, $\{(i,j)|i\in\mathcal{S}_u,j\in\mathcal{R}_u\}$, and $\{(v,j)|v\in\mathcal{V}_u,j\in\mathcal{R}_u\}$, with predefined probabilities $[\omega_{1}, \omega_{2}, \omega_{3}]$ respectively, where $\omega_{1}+\omega_{2}+\omega_{3}=1$. The generated training example is finally used to update the model parameters in \eqref{eq1}~(see \cite{BPR} for further details). 
We term the BPR method with this view-enhanced sampler as \textit{BPR+view$_{prob}$}.

Our proposed \textit{BPR+view$_{prob}$} uses biased sampling to exploit the side information provided by the viewed items. As each training example in \textit{BPR+view$_{prob}$} only contains two items, the viewed items are sampled as negative feedback and positive feedback with a probability of $\omega_1$ and $\omega_3$, respectively. In other words, it is hard to jointly learn the two-fold semantics of user preference on these viewed items, which assigns them a positive signal compared to those non-viewed items and a negative signal compared to purchased ones.
Therefore, next we move forward to improve BPR sampler by considering a view-enhanced weighted loss in objective function.
\subsubsection{Weighted Loss}                                                                                                                                                                                                                                                                                                                                                                                                           
To overcome the inefficacy issue in \textit{BPR+view$_{prob}$}, we propose to sample a item triple $(i,v,j)$ in each training example, where $i$, $v$ and $j$ represent a user's purchased item, viewed item and non-viewed item, respectively. Considering the user preference on these three items, the  model parameters, $\{\textbf{p}_u,\textbf{q}_i,\textbf{q}_v,\textbf{q}_j\}$, should be updated towards the objective of $\hat{r}_{ui} > \hat{r}_{uv} > \hat{r}_{uj}$. Therefore, similar to BPR, we design following objective function:
\begin{equation}
\begin{split}
	&J(\Theta)  =  \argmin_{\Theta} \sum_{(u,i,v,j)\in \mathcal{D}} -\ln \sigma(\hat{r}_{ui}(\Theta) - \hat{r}_{uj}(\Theta)) \\
	& - \alpha \ln \sigma(\hat{r}_{ui}(\Theta) - \hat{r}_{uv}(\Theta)) - (1-\alpha) \ln \sigma(\hat{r}_{uv}(\Theta) - \hat{r}_{uj}(\Theta)),
\end{split}
\label{eq2}
\end{equation}
where $\sigma(x)=1-\sigma(x)$, and $\Theta$ denotes the set of all parameters to be optimized. All three pairwise ranking relations among $i$, $v$ and $j$ are considered. Since the viewed item $v$ can be considered as both negative~($\hat{r}_{ui} > \hat{r}_{uv}$) and positive~($\hat{r}_{uv} > \hat{r}_{uj}$) feedback, the weighting parameter $\alpha$ in \eqref{eq2} controls the relative strength between these two semantics. Therefore, by tuning $\alpha$ empirically, we can train a model that properly exploits the user preference of view signal.

\begin{algorithm}[b!]
\caption{Learning Algorithm for \textit{BPR+view$_{loss}$}.}\label{alg1}
\SetKwInOut{Input}{Input}\SetKwInOut{Output}{Output}
\Input {purchase data $\mathcal{S}$, view data $\mathcal{V}$ }
\Output {$\Theta = \{ \textbf{P} \in \mathbb{R}^{M \times K}, \textbf{Q} \in \mathbb{R}^{N \times K}  \}$}
Randomly initialize \textbf{P} and \textbf{Q};\\
\While{not reaching convergence}{
// Random sampling\\
$u \gets$ draw a random user from $\mathcal{U}$\\
$i \gets$ draw a random purchased item from $\mathcal{S}_u$\\
$v \gets$ draw a random viewed item from $\mathcal{V}_u$\\
$j \gets$ draw a random non-viewed item from $\mathcal{R}_u$\\
// Eq.~\eqref{eq5} - \eqref{eq8}\\
Compute gradients of $\{\textbf{p}_u,\textbf{q}_i,\textbf{q}_v,\textbf{q}_j\}$\\
// Eq.~\eqref{eq4}\\
Update the above parameters
}
\end{algorithm} 

Note that we have omitted $L_2$ regularization terms for clarity. We use matrix factorization to predict $\hat{r}_{ui}$, user $u$'s preference on item $i$, obtained by calculating the dot product of the latent factors of the user $\textbf{p}_u$ and the item $\textbf{q}_i$, as follows:
\begin{equation}
	\hat{r}_{ui} = \textbf{p}^T_u \textbf{q}_i = \sum_{f=1}^K p_{u,f} \times q_{i,f}.
\label{eq3}
\end{equation}
Recall that $K$ is the number of latent factors. Finally, we use Stochastic Gradient Descent~(SGD) to find a local minimum of the objective function in \eqref{eq2}. In particular, for each iteration (Algorithm~\ref{alg1}, Lines: 3-11), given a random feedback triple of user $u$ who has purchased item $i$, viewed~(but not purchased) item $v$ but not viewed item $j$, $(u,i,v,j)\in \mathcal{D}= \{ (u,i,v,j)|i \in \mathcal{S}_u \land v \in \mathcal{V}_u \land j \in \mathcal{R}_u \}$, we update the model parameter $\theta \in \Theta$ based on the gradient of its corresponding parameter $\frac{\partial J}{\partial \theta}$ while fixing the others, until convergence, as follows:
\begin{equation}
	\theta^{(t+1)} = \theta^{(t)} + \eta^{(t)} \cdot \frac{\partial J}{\partial \theta}(\theta^{(t)}).
\label{eq4}
\end{equation}
Note that learning rate parameter $\eta$ can both be a fixed constant or an adaptive value like Adagrad~\cite{Adagrad}. The gradients of latent vectors $\{\textbf{p}_u,\textbf{q}_i,\textbf{q}_v,\textbf{q}_j\}$ are calculated as follows:
\begin{equation}
\begin{split}
	\frac{\partial J}{\partial \textbf{p}_u} =\ &\delta(\hat{r}_{ui} - \hat{r}_{uj}) (\textbf{q}_i-\textbf{q}_j) + \alpha \delta(\hat{r}_{ui} - \hat{r}_{uv}) (\textbf{q}_i-\textbf{q}_v) \\
	&+ (1-\alpha) \delta(\hat{r}_{uv} - \hat{r}_{uj}) (\textbf{q}_v-\textbf{q}_j) - \lambda \textbf{p}_u,
\end{split}
\label{eq5}
\end{equation}
\begin{equation}
\begin{split}
	\frac{\partial J}{\partial \textbf{q}_i} =\ &\delta(\hat{r}_{ui} - \hat{r}_{uj}) \textbf{p}_u + \alpha \delta(\hat{r}_{ui} - \hat{r}_{uv}) \textbf{p}_u \\
	&+ (1-\alpha) \delta(\hat{r}_{uv} - \hat{r}_{uj}) \textbf{p}_u - \lambda \textbf{q}_i,
\end{split}
\label{eq6}
\end{equation}
\begin{equation}
	\frac{\partial J}{\partial \textbf{q}_v} = - \alpha \delta(\hat{r}_{ui} - \hat{r}_{uv}) \textbf{p}_u + (1-\alpha) \delta(\hat{r}_{uv} - \hat{r}_{uj}) \textbf{p}_u - \lambda \textbf{q}_v,
\label{eq7}
\end{equation}
\begin{equation}
	\frac{\partial J}{\partial \textbf{q}_j} = - \delta(\hat{r}_{ui} - \hat{r}_{uj}) \textbf{p}_u - (1-\alpha) \delta(\hat{r}_{uv} - \hat{r}_{uj}) \textbf{p}_u - \lambda \textbf{q}_j,
\label{eq8}
\end{equation}
 where the regularization parameter $\lambda$ is added to avoid overfitting. Regarding the complexity of the above pairwise learning algorithm, the computation of each gradient is $O(K)$, where $K$ is the number of latent factors. The total complexity is $O(T \cdot K)$, where $T$ is the number of iterations.
We term the above variant of BPR sampler as \textit{BPR+view$_{loss}$}.

\begin{figure*}[t]
\centering
\subfloat[Beibei]{\includegraphics[width=.320\textwidth]{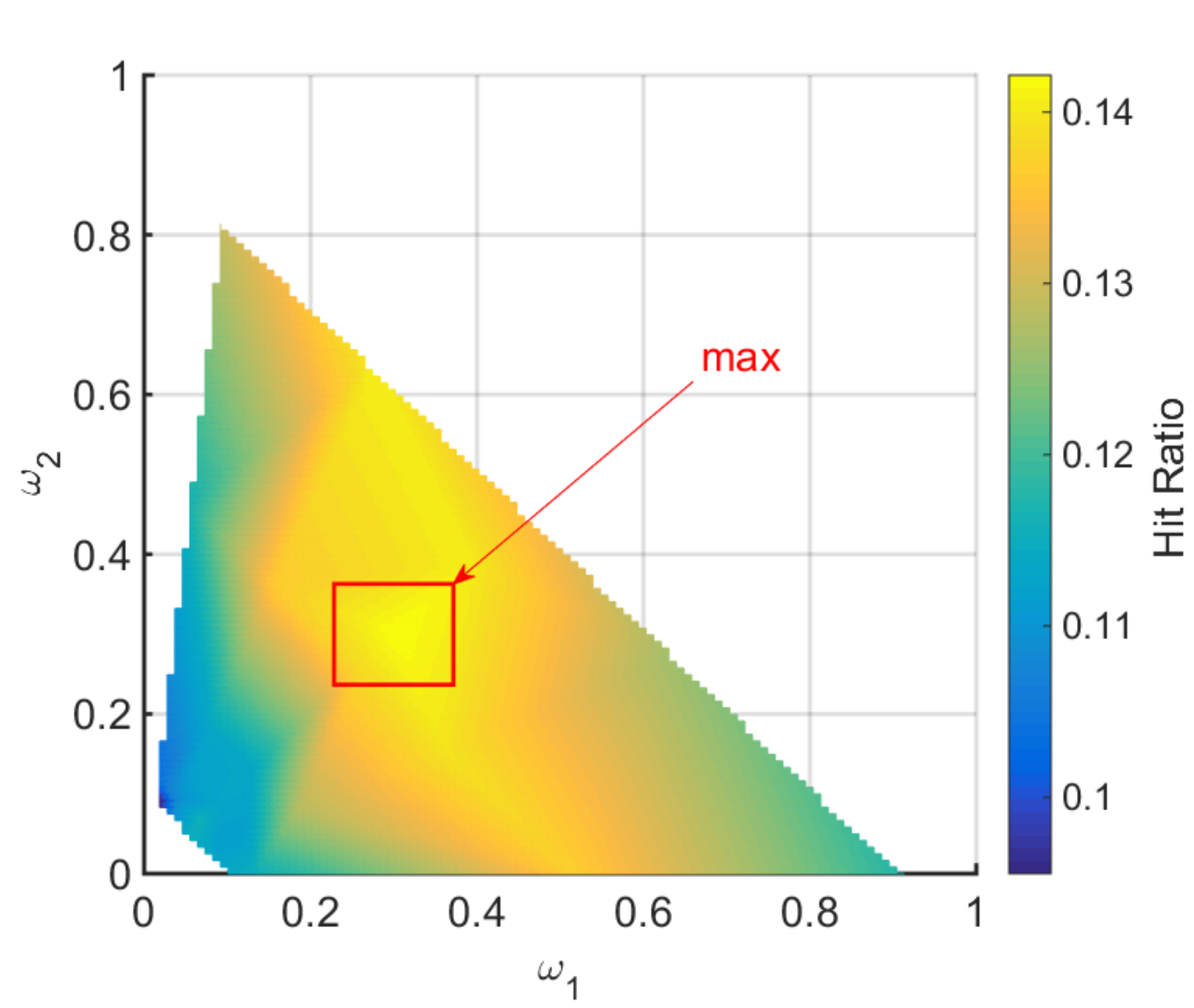}}
\hfill
\subfloat[Tmall-all]{\includegraphics[width=.320\textwidth]{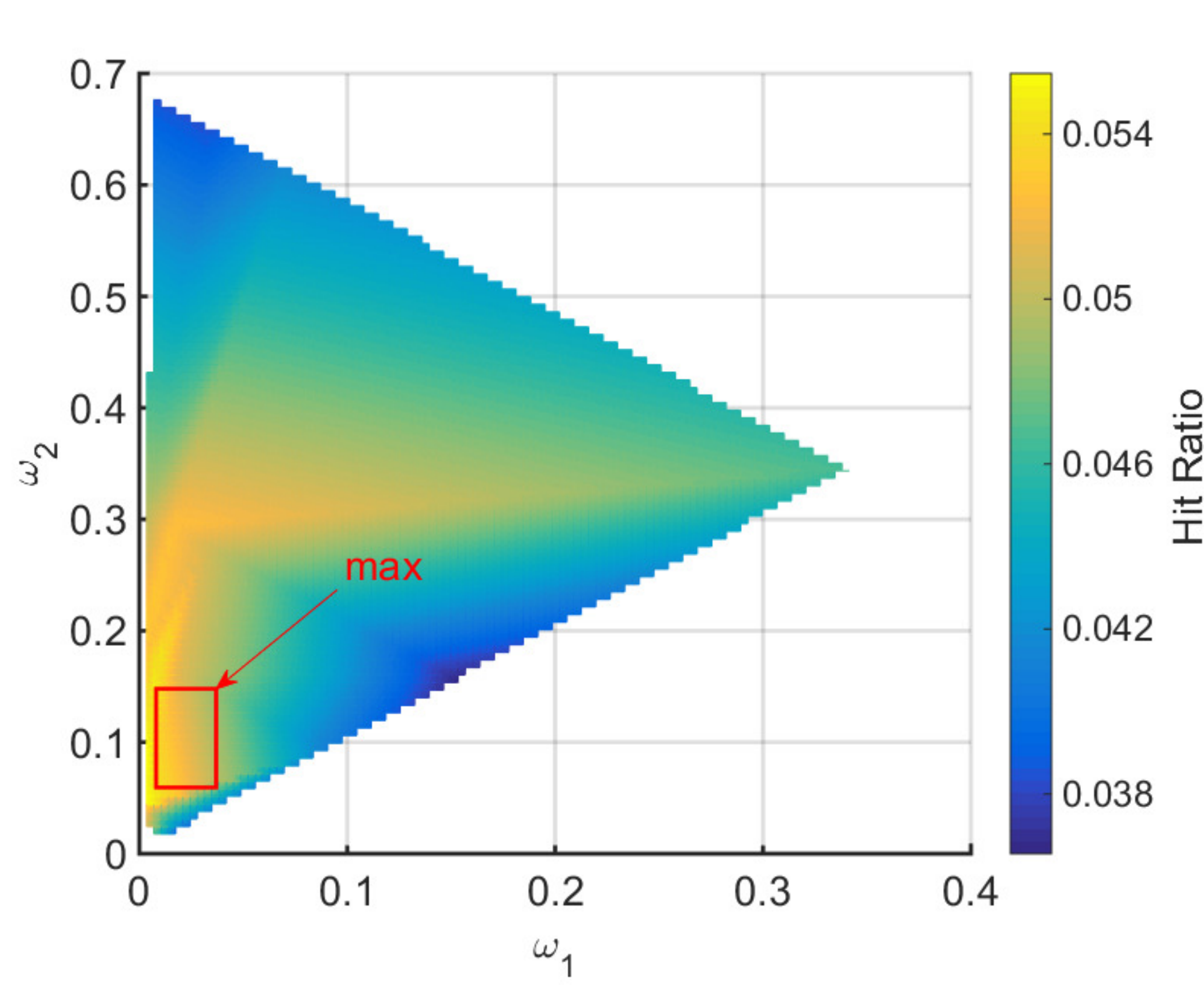}}
\hfill
\subfloat[Tmall-selected]{\includegraphics[width=.320\textwidth]{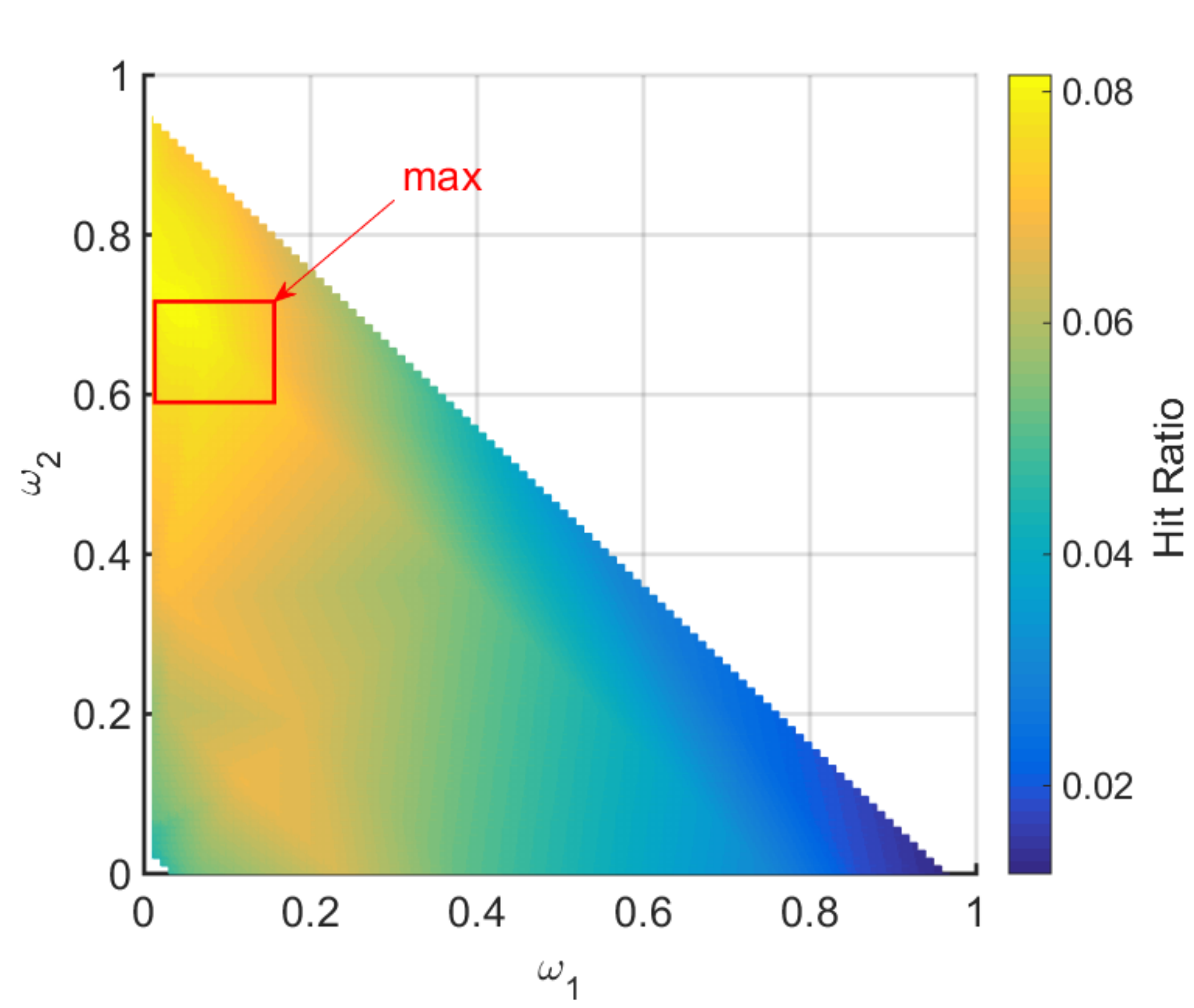}}
\caption{Impact of sampling probability parameters $\{\omega_1,\omega_2,\omega_3\}$ on \textit{BPR+view$_{prob}$}'s performance, in terms of HR.}
\label{fig_w_explore}
\end{figure*}

\begin{figure*}[t]
\centering
\subfloat[\textit{BPR+view$_{loss}$}, Beibei]{\includegraphics[width=.320\textwidth]{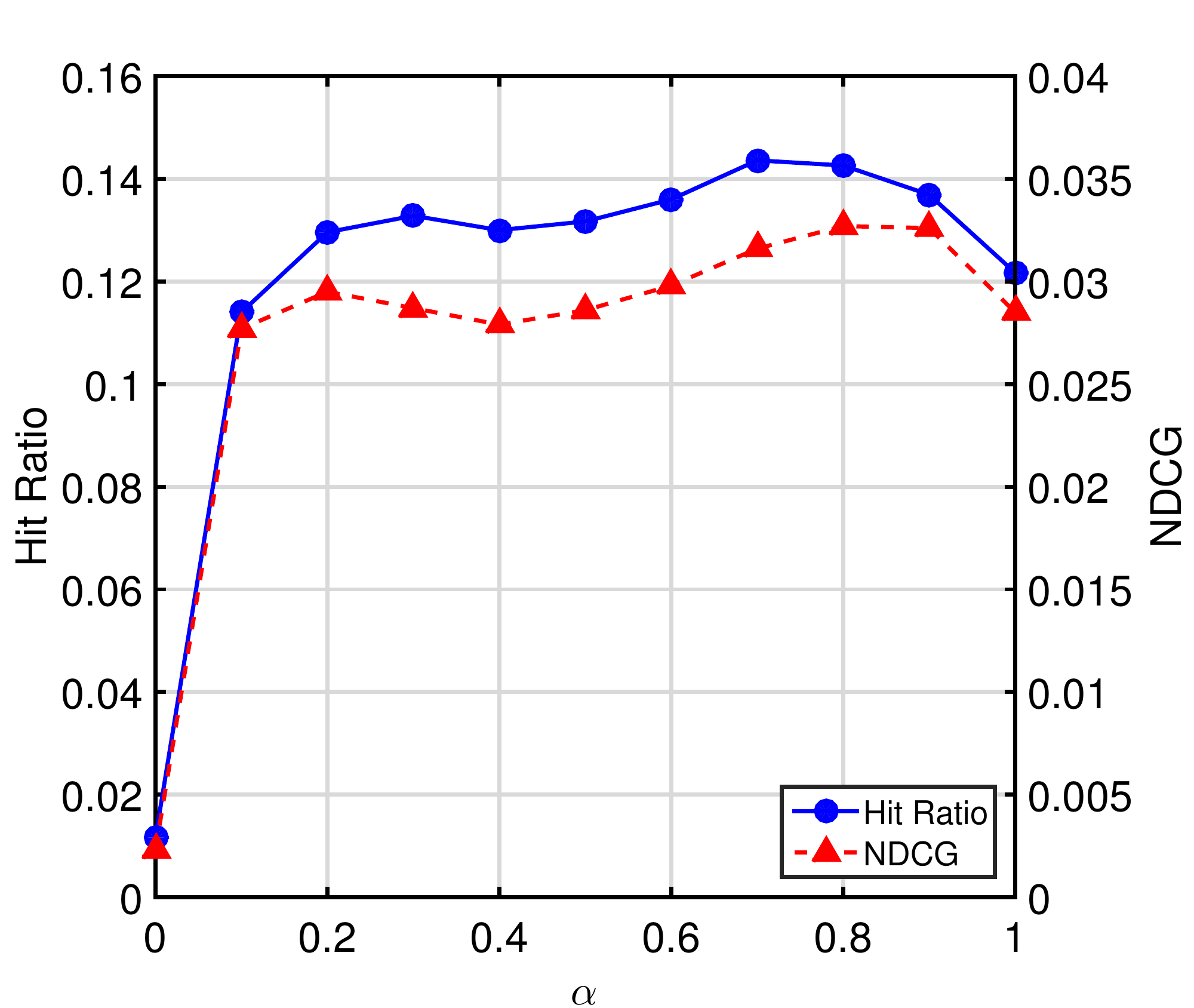}}
\hfill
\subfloat[\textit{BPR+view$_{loss}$}, Tmall-selected]{\includegraphics[width=.320\textwidth]{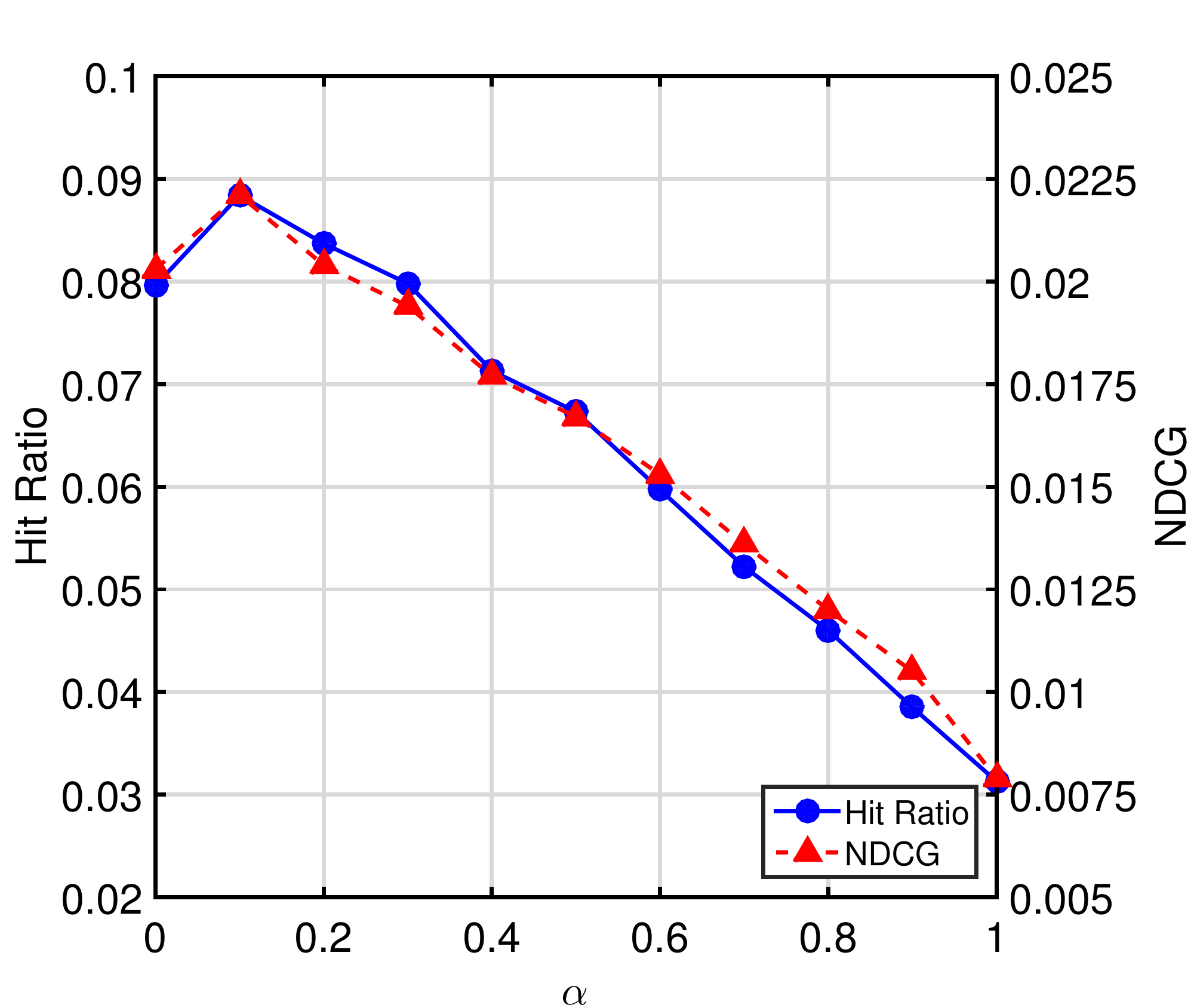}}
\hfill
\subfloat[\textit{BPR+view$_{loss}^{\beta}$}, Beibei]{\includegraphics[width=.320\textwidth]{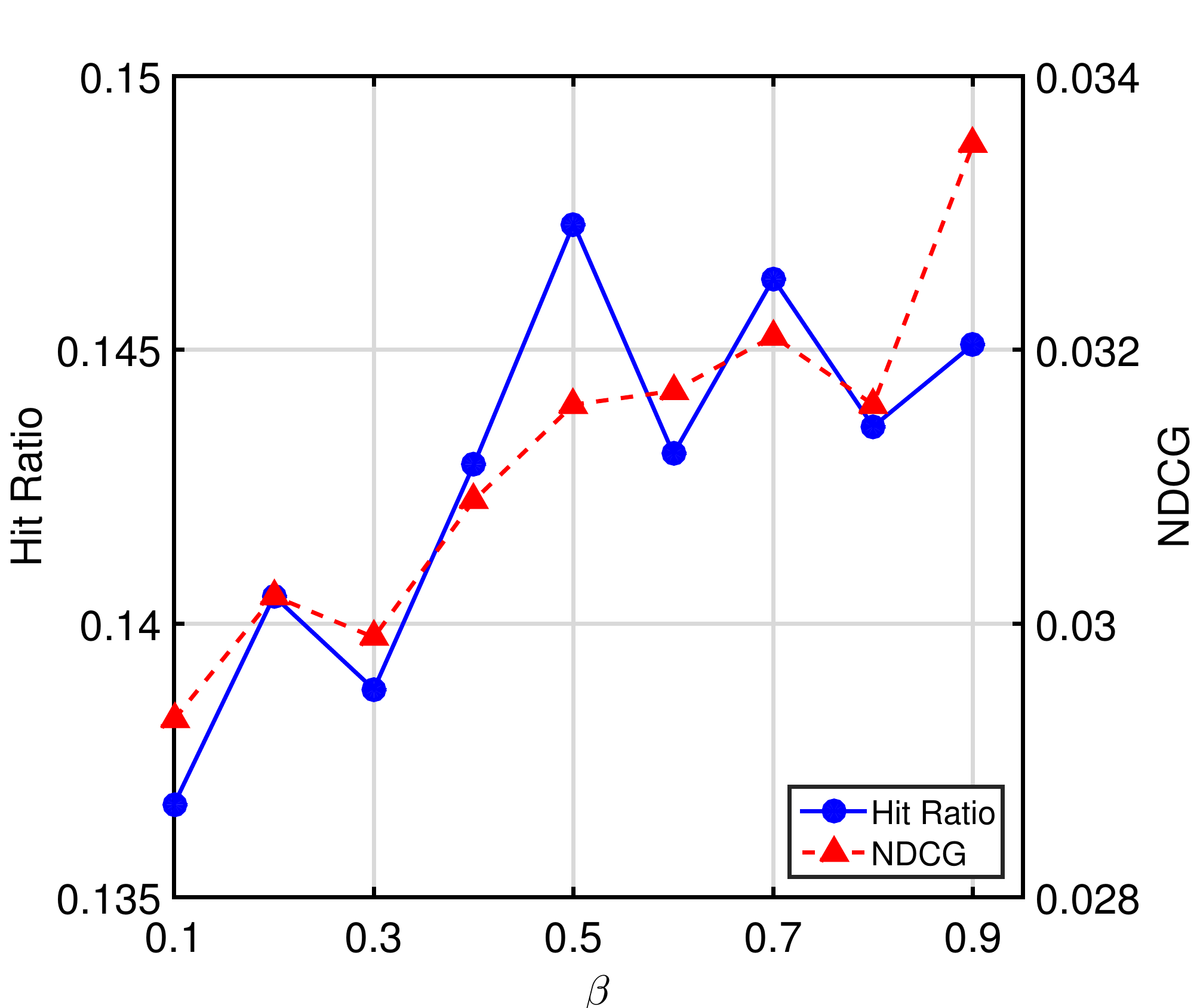}}
\caption{Impact of weighting parameters $\alpha$ and $\beta$ on HR performances of \textit{BPR+view$_{loss}$} and \textit{BPR+view$_{loss}^{\beta}$}, respectively.}
\label{fig_alpha_explore}
\end{figure*}

\subsection{User-aware Weighting Strategy}
%In our collected data, the number of viewed interactions are generally 5 to 10 times more than that of purchased ones, indicating that a user may view many items before she finally decides which to buy. 
Intuitively, if a user tend to view many items and instead purchase another one, the viewed interactions should indicate a stronger negative signal than that of other users. In this meaning, the relative strength between two semantics of view signal should differ among users. Let $A_u$ denote a user $u$'s view-purchase ratio that measures the degree of whether $u$ prefers to view many items before deciding which to buy, it is reasonable to think a higher $A_u$ indicates a stronger negative signal in $u$'s viewed interactions, which corresponds to a higher weight $\alpha$ in our proposed \textit{BPR+view$_{loss}$}. To account for this effect, we parametrize a user-oriented weight $\alpha_u$ based on $A_u$:
\begin{equation}
	\alpha_u = \frac{A^{\beta}_u}{(A^{\beta}_u+1)},
\label{eq9}
\end{equation}
where exponent $\beta$ controls the significance level of this effect --- it is strengthened when $\beta>1$ and smoothed while setting $\beta \in (0,1)$. We term this new BPR sampler with user-aware weighting scheme as \textit{BPR+view$^{\beta}_{loss}$}

Next, we focus on the definition of view-purchase ratio $A_u$ above. A straightforward way of computing it would be the ratio between number of user $u$'s viewed interactions and purchased ones. However, as users' shopping history is divided into several sessions, computing $A_u$ in the session-level can be more accurate. More specifically, we define $A_u$ as the average value among these sessions:
\begin{equation}
	A_u = \frac{\sum_{s=1}^S{a_{u,s}}}{|S|},\ a_{u,s} = \frac{\mathcal{V}_{u,s}}{\mathcal{P}_{u,s}},
\label{eq10}
\end{equation}
where $a_{u,s}$, $\mathcal{V}_{u,s}$ and $\mathcal{R}_{u,s}$ represent $u$'s view-purchase ratio, viewed item set and purchased item set in session $s$, respectively. To generate $u$'s sessions in the shopping history, we first sort $u$'s viewed and purchased interactions according to timestamps and then we merge those consecutive interactions into one session based on whether they happen within a threshold $d$. Since the suitable setting of $d$ may vary between different datasets, we empirically tune this parameter and search the best recommendation performance. The result shows that $d=3600$~(s) works well in Beibei dataset. 
As for Tmall dataset, since the timestamp information only contains the date, it is infeasible to extract session information in each user's shopping history. Therefore, we leave the exploration of user-aware weighting scheme on \textit{BPR+view$^{\beta}_{loss}$} for future work.

\subsection{Results}
We first study the influence of hyper-parameters. Then we compare the performance of our proposed BPR sampler with the original one.

\subsubsection{Hyper-parameter Investigation}\
\mypara{\textit{BPR+view$_{prob}$}.}
In the biased sampling, our proposed \textit{BPR+view$_{prob}$} has three non-negative parameters: $[\omega_1,\omega_2,\omega_3]$, which respectively represents the probability of item pairs among users' purchased, viewed and unobserved interactions. Considering $\omega_{1}+\omega_{2}+\omega_{3}=1$, we have to search two independent parameters. 
Fig.~\ref{fig_w_explore} shows its performance~(HR) with different $\omega_{1}$ and $\omega_{2}$. On Beibei dataset, \textit{BPR+view$_{prob}$} performs best when $[\omega_1,\omega_2,\omega_3]=[0.3,0.3,0.4]$ , as shown by the yellow center of Fig.~\ref{fig_w_explore}(a). In terms of the two-fold semantics encoded in view data, we use $\frac{\omega_{3}}{\omega_{1}}$ to measure whether it is more closed to positive feedback~($>1$) or negative feedback~($<1$). Here in Beibei, $\frac{\omega_{3}}{\omega_{1}}$ is close to $1$, indicating both two folds are equally important.
However, in Tmall-all dataset, the best performance appears when $[\omega_1,\omega_2,\omega_3]= [0.01,0.09,0.9]$, indicating that view data acts as a strong indicator of user preference, more important than purchase data as $\omega_2<\omega_3$. Recall that the test data in Tmall-all belongs to Nov. 11$^{th}$, an annual global shopping festival on Tmall platform, we believe this abnormal observation of view data was caused by the fact that most users had viewed a lot of items in their wish lists before Nov. 11$^{th}$ and these viewed items indicated a strong positive signal even though they did not purchased them in the end.
Thus we also investigate $[\omega_1,\omega_2,\omega_3]$ in a subset Tmall data, Tmall-selected, where the above effect is avoided~(See Section~\ref{sec:dataset} for more details). In Tmall-selected, as shown in Fig.~\ref{fig_w_explore}(c), peak performance lies in [0.01,0.74,0.25]. Unlike Tmall-all, view data is less important than purchase data, with $\omega_2>\omega_3$. When compared with that in Beibei dataset, view data in Tmall-selected is still more close to a positive feedback due to larger value of $\frac{\omega_{3}}{\omega_{1}}$. Because of abnormally intensive influence of viewed interactions, following experiments will not take Tmall-all into consideration.

\mypara{\textit{BPR+view$_{loss}$}.}
Now, we study the impact of weighting parameter $\alpha$ on \textit{BPR+view$_{loss}$}. 
As shown in Fig.~\ref{fig_alpha_explore}(a), we observe the best $\alpha$ varies between 0.7 and 0.8 on Beibei. Since a large $\alpha$ increases the importance of learning user preference from purchased and viewed item pairs, this observation highlights the significance of considering users' viewing behavoirs more as a negative feedback. However, the performance still shows a drop at $\alpha=1$, where we take viewed items as equally important as those unobserved ones. This observation also confirms the necessity of taking view interactions as a weak positive feedback. In Fig.~\ref{fig_alpha_explore}(b), the performance drop steeply as $\alpha$ increases in Tmall-selected dataset and the peak lies at $\alpha=0.1$, where view items are almost utilized equally as purchased ones but pairwise ranking relation between them still exists. The performance of \textit{BPR+view$_{loss}$} is sensitive to $\alpha$ in Tmall-selected, while not in Beibei. This difference may be caused by the same reason as distinctive influence of $[\omega_1,\omega_2,\omega_3]$ on \textit{BPR+view$_{prob}$} mentioned above, that view data represents a more effective signal of user preference in Tmall-selected dataset. 

\begin{table}[b]
\small
\renewcommand{\arraystretch}{1.2}
\setlength\tabcolsep{4.0pt}
\centering
\caption{Performance comparison among \textit{BPR}, \textit{BPR+view$_{prob}$}, \textit{BPR+view$_{loss}$} and \textit{BPR+view$^{\beta}_{loss}$}.}
\label{tab_perf_comp}
\centering
\subfloat[Beibei]{
\begin{tabular}{|c|c|c|c|c|}
\hline
                                   & \textbf{HR} & \textbf{$\Delta$} & \textbf{NDCG} & \textbf{$\Delta$} \\ \hline
\textbf{BPR~(baseline)}       & 0.1086      & --                & 0.0242        & --                \\ \hline
\textbf{BPR+view$_{prob}$}         & 0.1422      & +30.93\%          & 0.0321        & +32.64\%          \\ \hline
\textbf{BPR+view$_{loss}$}         & 0.1436      & +32.23\%          & 0.0327        & +35.12\%          \\ \hline
\textbf{BPR+view$^{\beta}_{loss}$} & 0.1473      & +35.64\%          & 0.0335        & +38.43\%          \\ \hline
\end{tabular}
\label{tab:beibei}
}
\vfill
\subfloat[Tmall-selected]{
\begin{tabular}{|c|c|c|c|c|}
\hline
                             & \textbf{HR} & \textbf{$\Delta$} & \textbf{NDCG} & \textbf{$\Delta$} \\ \hline
\textbf{BPR~(baseline)} & 0.0755      & --                & 0.0191        & --                \\ \hline
\textbf{BPR+view$_{prob}$}   & 0.0807      & +6.89\%           & 0.0199        & +4.19\%           \\ \hline
\textbf{BPR+view$_{loss}$}   & 0.0884       & +17.09\%          & 0.0221        & +15.71\%          \\ \hline
\end{tabular}
\label{tab:Tmall-selected}
}
\end{table}

\mypara{\textit{BPR+view$^{\beta}_{loss}$}.}
Fig.~\ref{fig_alpha_explore}(c) plots the prediction accuracy of \textit{BPR+view$^{\beta}_{loss}$} on Beibei, with different $\beta$. This model achieves best performance at ${\beta}=0.5$ evaluated by HR and $0.9$ by NDCG. Such $\beta$ less than 1 smooths the effect of $A_u$ on $\alpha_u$, indicating a weak but still influential relationship between $A$ and the confidence of view signal. As for Tmall-selected, since we cannot extract users' shopping sessions from the coarse-grained timestamp in each record, we do not conduct similar experiments on this dataset.  

According to the investigation above, we fix these hyper-parameters according to the best performance evaluated by HR, i.e., $[\omega_1,\omega_2,\omega_3]=[0.3,0.3,0.4],\alpha=0.7,\beta=0.5$ for Beibei and $[\omega_1,\omega_2,\omega_3]=[0.01,0.74,0.25],\alpha=0.1$ for Tmall-selected.

\subsubsection{Performance Comparison}
We compare the performance of vanilla \textit{BPR} and our proposed view-enhanced sampler. The main result is listed in Table~\ref{tab_perf_comp}.

\begin{figure}[t]
\centering
\subfloat[HR, Beibei]{\includegraphics[width=.255\textwidth]{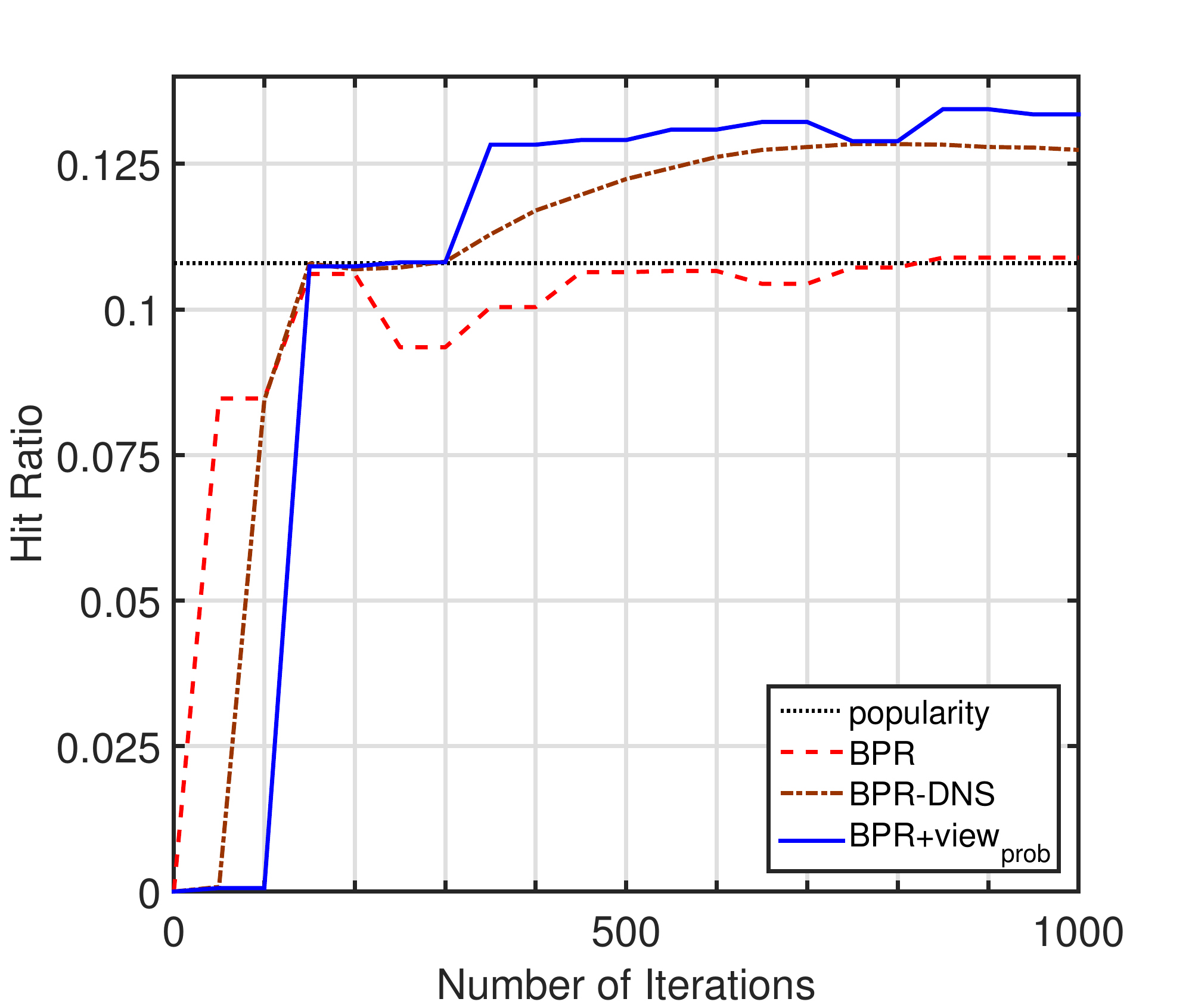}}
\hfill\hspace{-2em}
\subfloat[NDCG, Beibei]{\includegraphics[width=.255\textwidth]{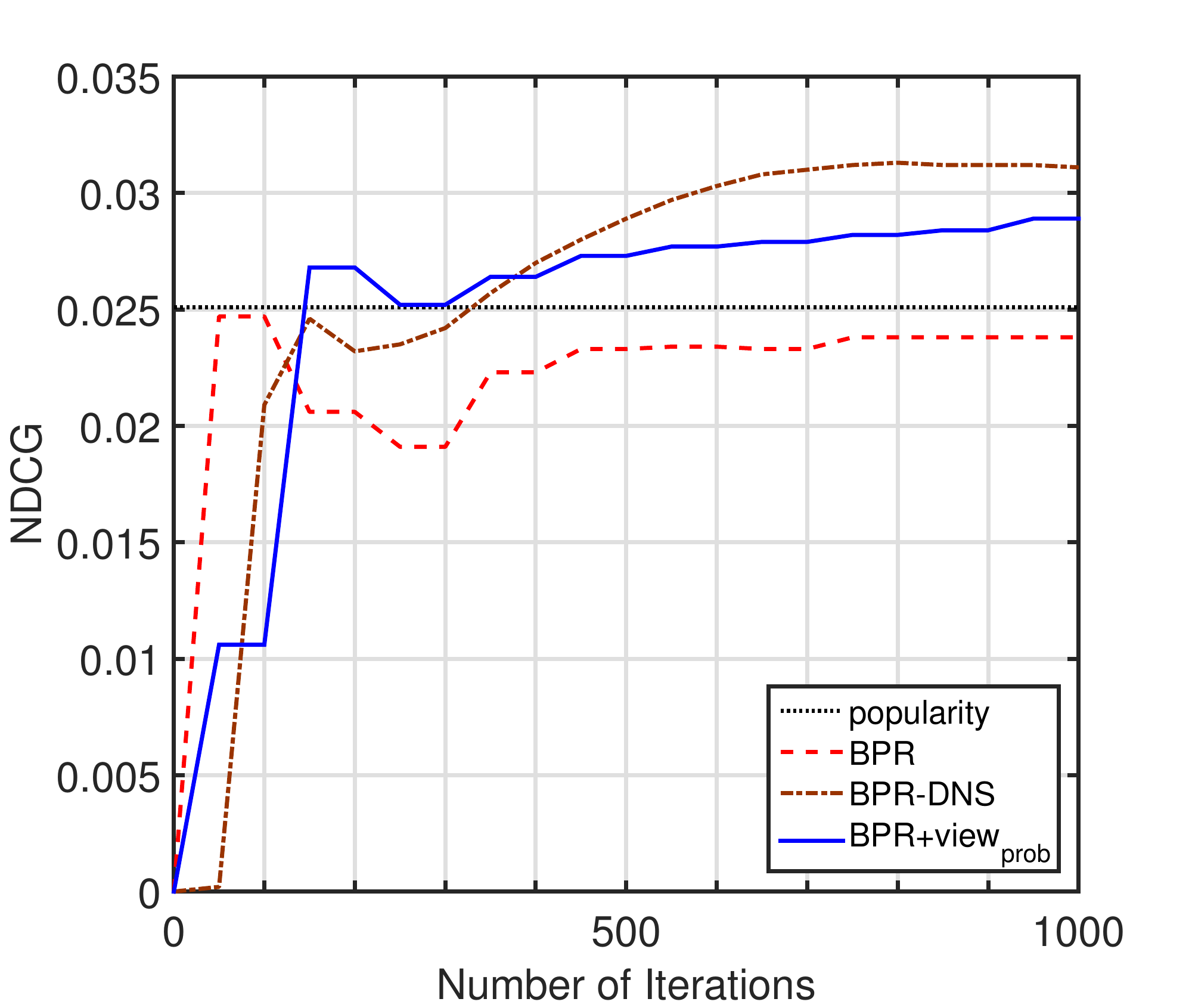}}
\vfill
\subfloat[HR, Tmall-all]{\includegraphics[width=.255\textwidth]{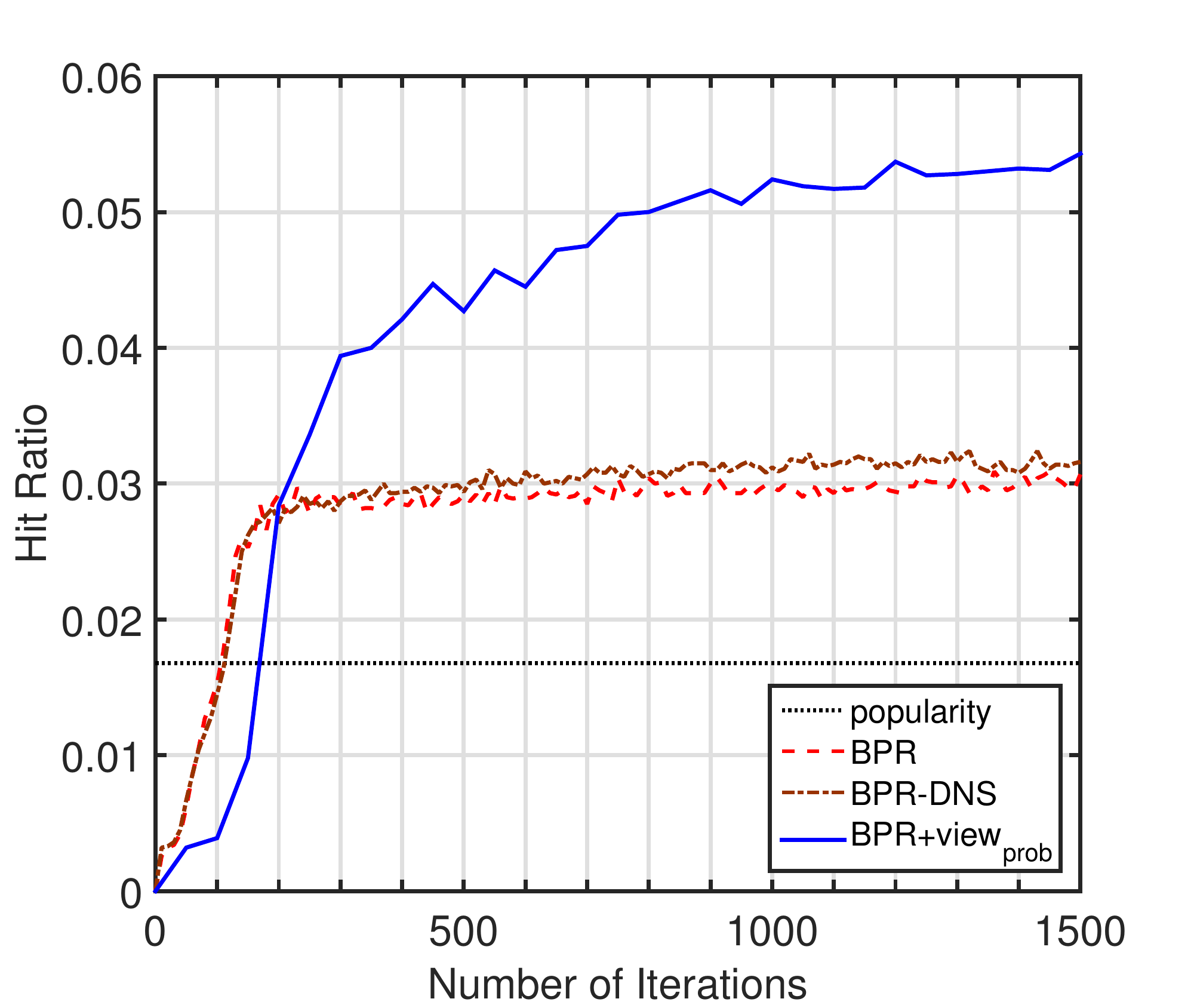}}
\hfill\hspace{-2em}
\subfloat[NDCG, Tmall-all]{\includegraphics[width=.255\textwidth]{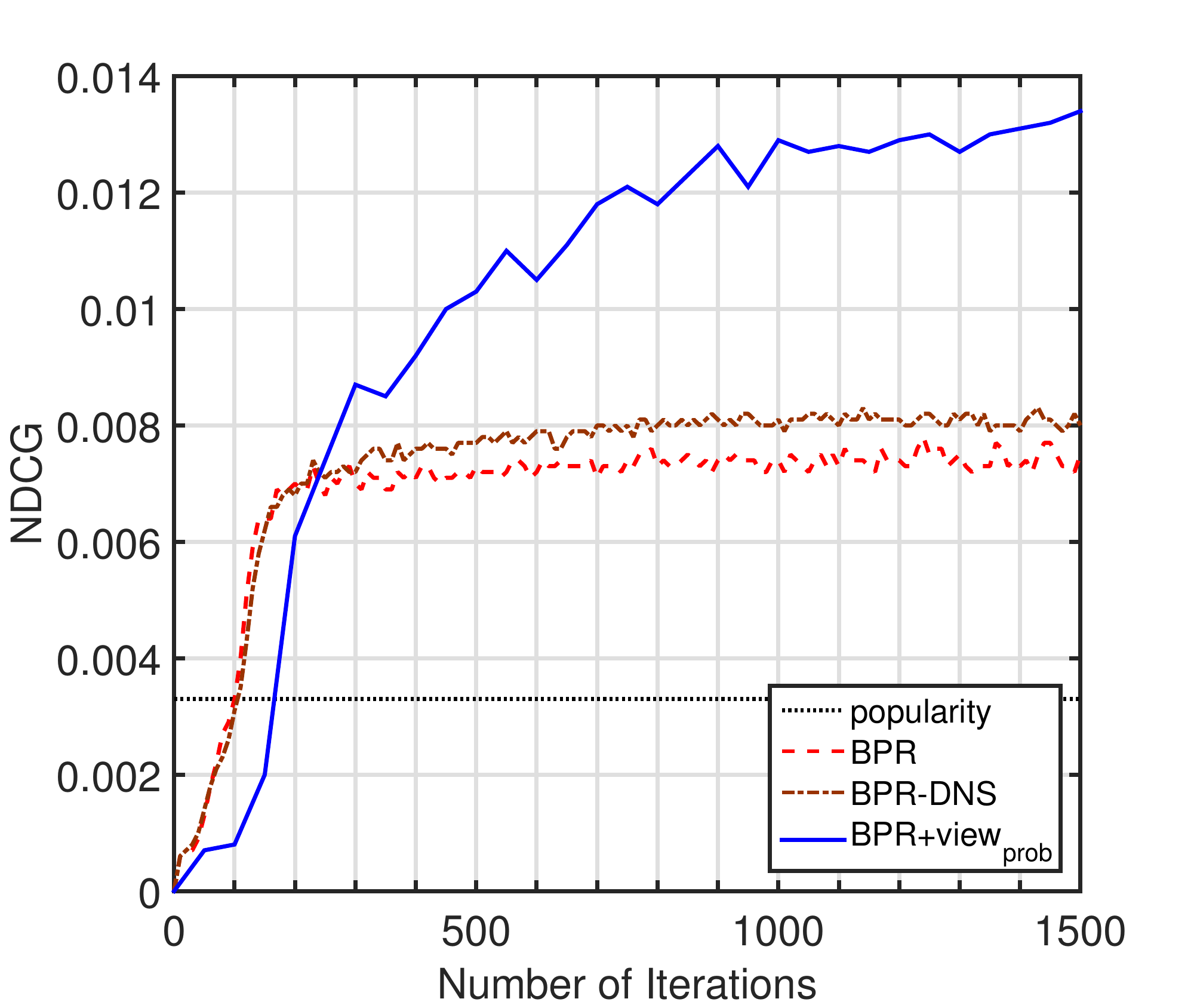}}
\vfill
\subfloat[NDCG, Tmall-select]{\includegraphics[width=.255\textwidth]{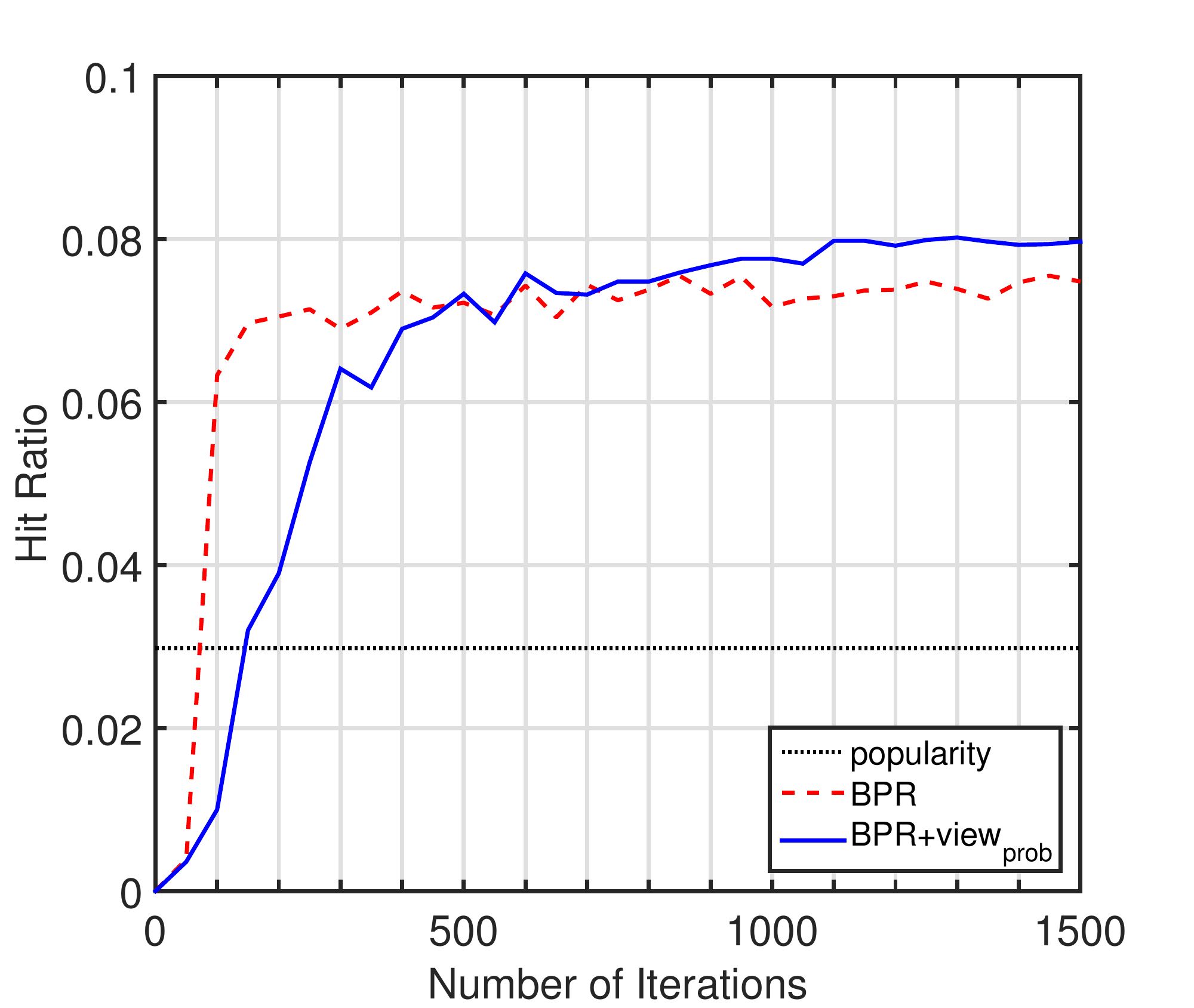}}
\hfill\hspace{-2em}
\subfloat[NDCG, Tmall-selected]{\includegraphics[width=.255\textwidth]{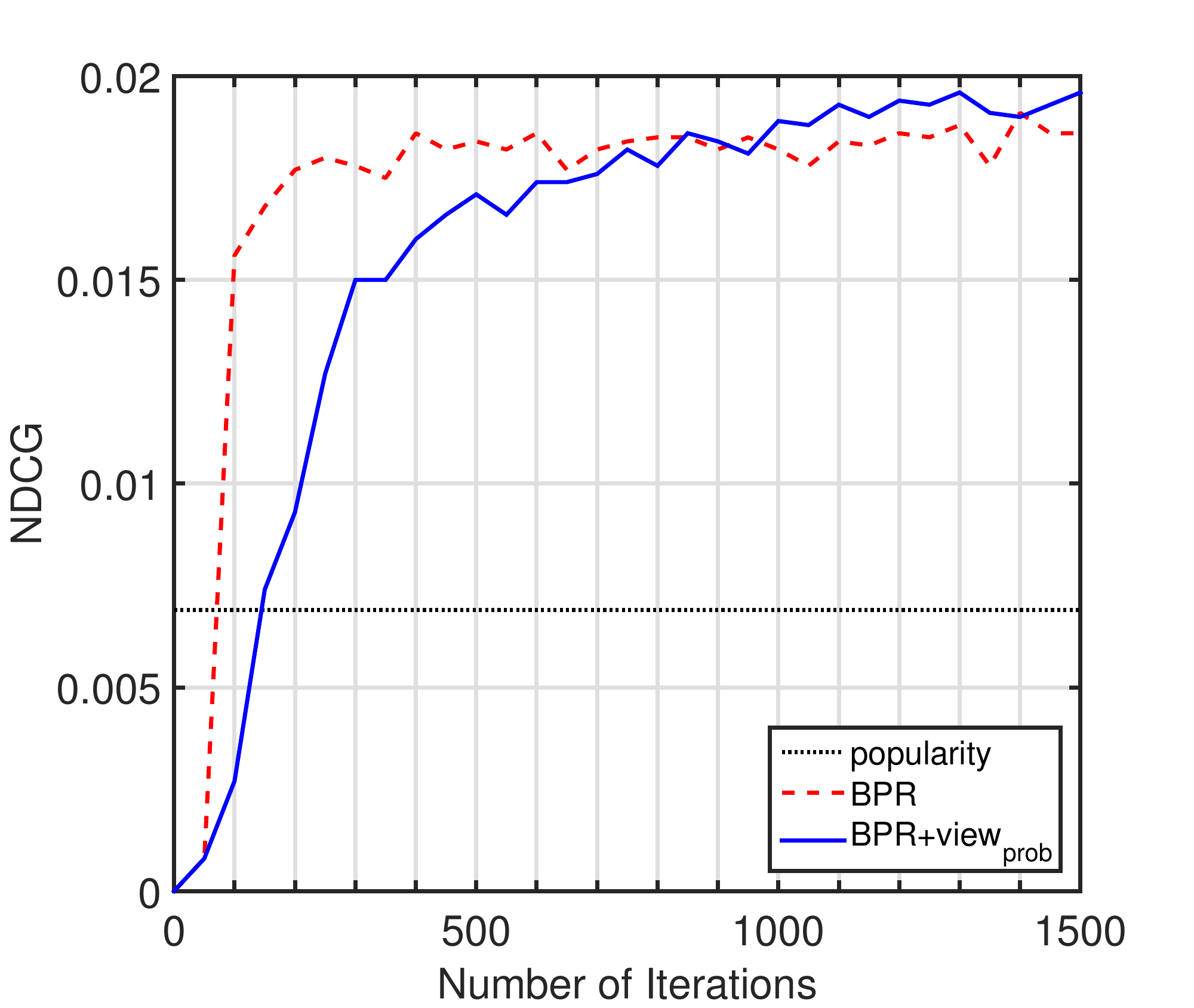}}
\caption{Performance comparison in each iteration~(\textit{BPR+view$_{prob}$}).}
\label{fig_perf_comp_1}
\end{figure}

\mypara{\textit{BPR+view$_{prob}$} vs. \textit{BPR}.}
Our proposed \textit{BPR+view} achieves the best performance when $[\omega_{1}, \omega_{2}, \omega_{3}]$ are set as $[0.3, 0.3, 0.4]$ and $[0.01, 0.09, 0.9]$ on the Beibei and Tmall datasets, respectively. 
To demonstrate its effectiveness, we compare it with 1) the vanilla \textit{BPR}~\cite{BPR}, and 2) \textit{BPR-DNS}~\cite{DNS}, which selects the item with the highest prediction score among $X$ randomly sampled negatives. 
For \textit{BPR-DNS}, we tune the $X$ in the same way as the original paper. 
To our knowledge, DNS is the most effective sampler to date for BPR based on the interaction data only, and empirically outperforms \cite{rendle2014improving}. 
In addition, we evaluate a common baseline \textit{Popularity}~\cite{NCF}, which simply recommends items based on their popularity evidenced by the number of purchases.

Fig.~\ref{fig_perf_comp_1} shows the testing HR and NDCG of the compared methods in each training iteration. As can be seen, upon convergence, \textit{BPR+view} significantly outperforms all other methods on three datasets, except for the NDCG on Beibei. This justifies the efficacy of accounting for the preference signal in the view data using our proposed sampler. 
Besides, the relative improvements of \textit{BPR+view} over \textit{BPR} are about 30\%+, 80\%+ and 5\%+ on Beibei, Tmall-all and Tmall-selected dataset, respectively~(See Table~\ref{tab_perf_comp}). Due to the effect of Global Shopping Festival on Tmall-all, the improvement is more significant. Last but not least, we observed that \textit{Popularity} performs as well as \textit{BPR} on the Beibei dataset, which is unexpected since \textit{BPR} is a personalized recommendation method. Our further investigation finds that it is because the Beibei dataset is highly popularity-skewed --- the top-1\% items contribute almost 50\% of purchases, as illustrated in Fig.~\ref{fig_skewness}(a).

Clearly, after integrating viewing signal as intermediate feedback, \textit{BPR+view$_{prob}$} outperforms the original \textit{BPR} that only contains purchase feedback.

\begin{figure}[t]
\centering
\subfloat[HR, Beibei]{\includegraphics[width=.255\textwidth]{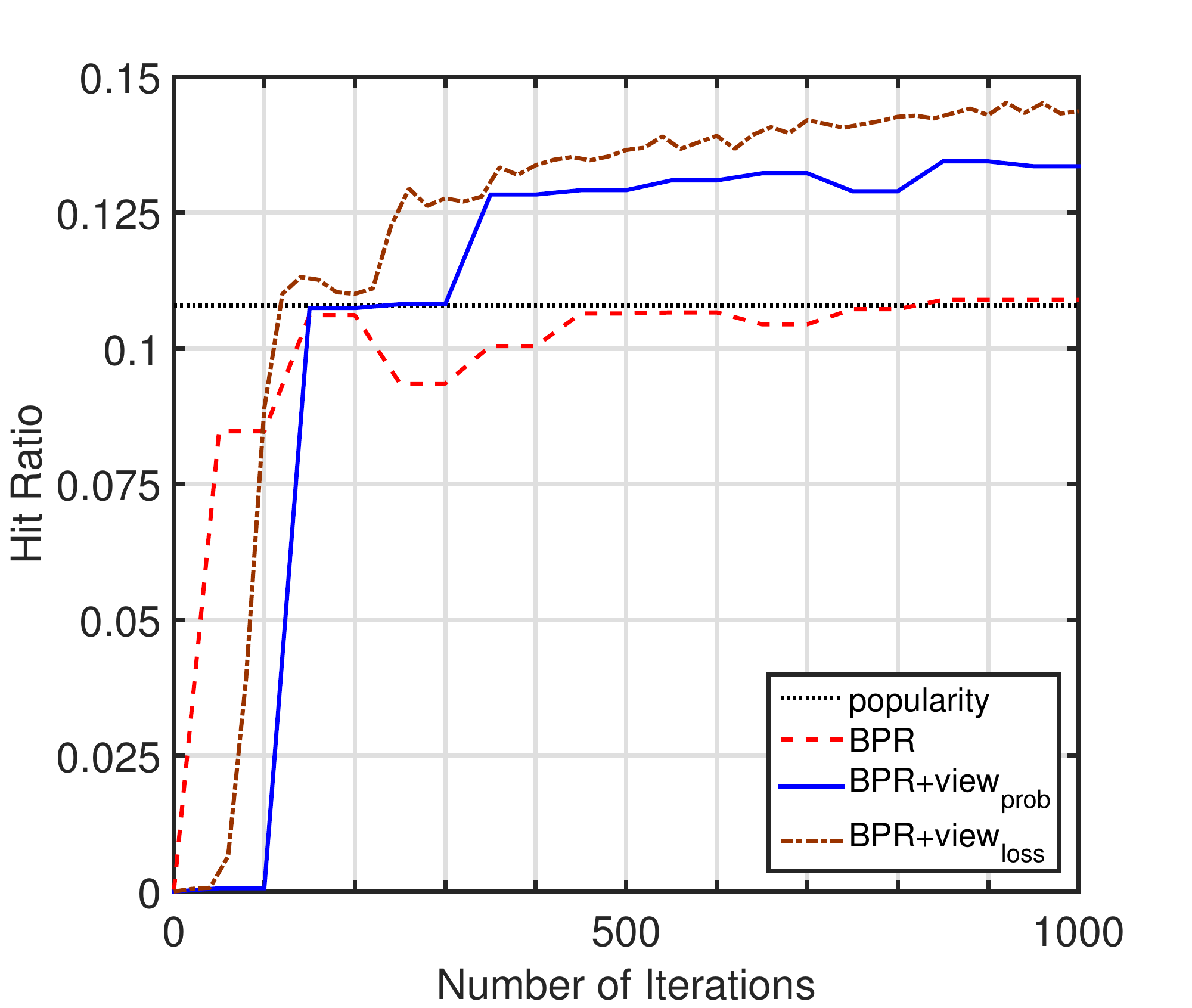}}
\hfill\hspace{-2em}
\subfloat[NDCG, Beibei]{\includegraphics[width=.255\textwidth]{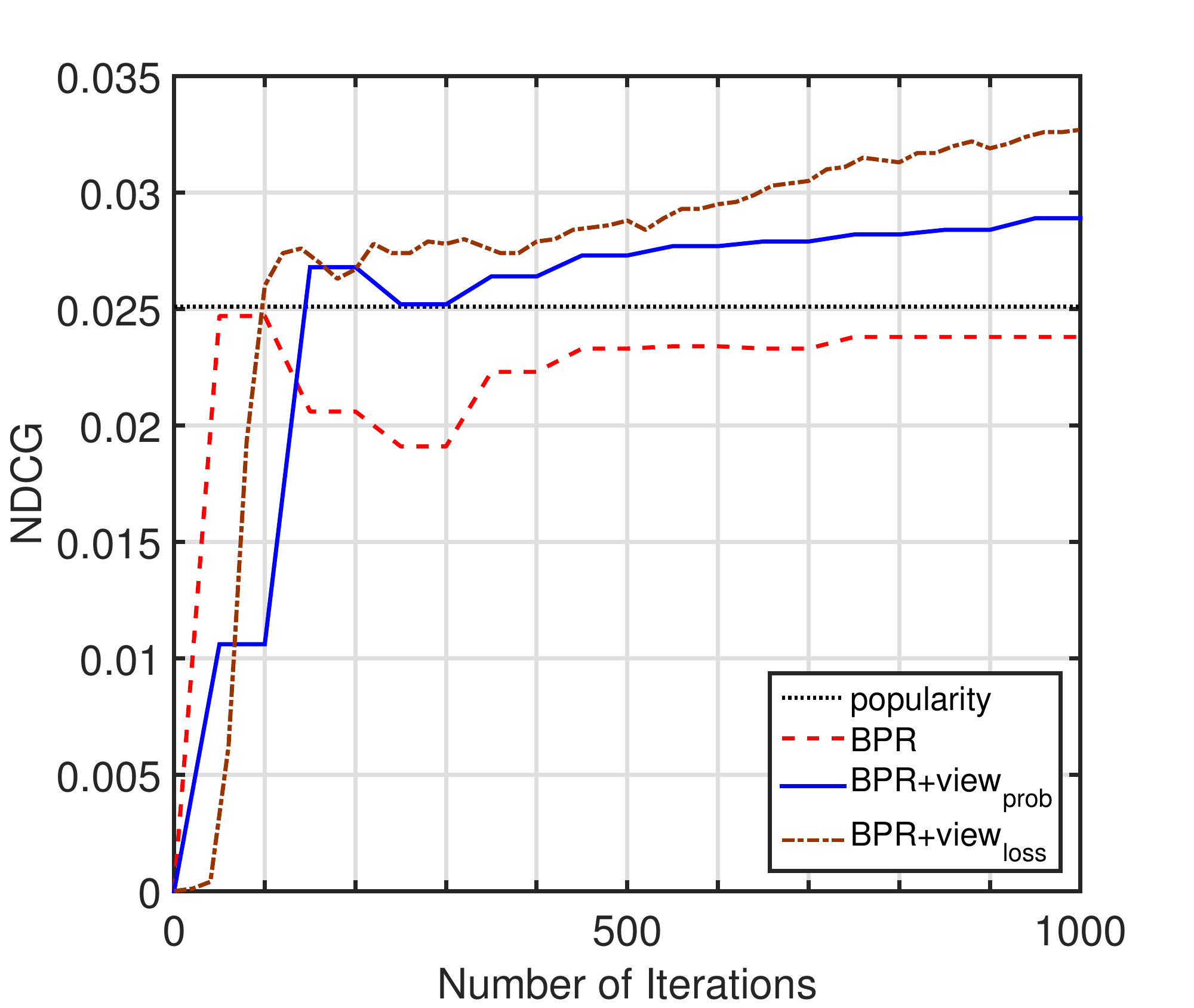}}
\vfill
\subfloat[HR, Tmall-selected]{\includegraphics[width=.255\textwidth]{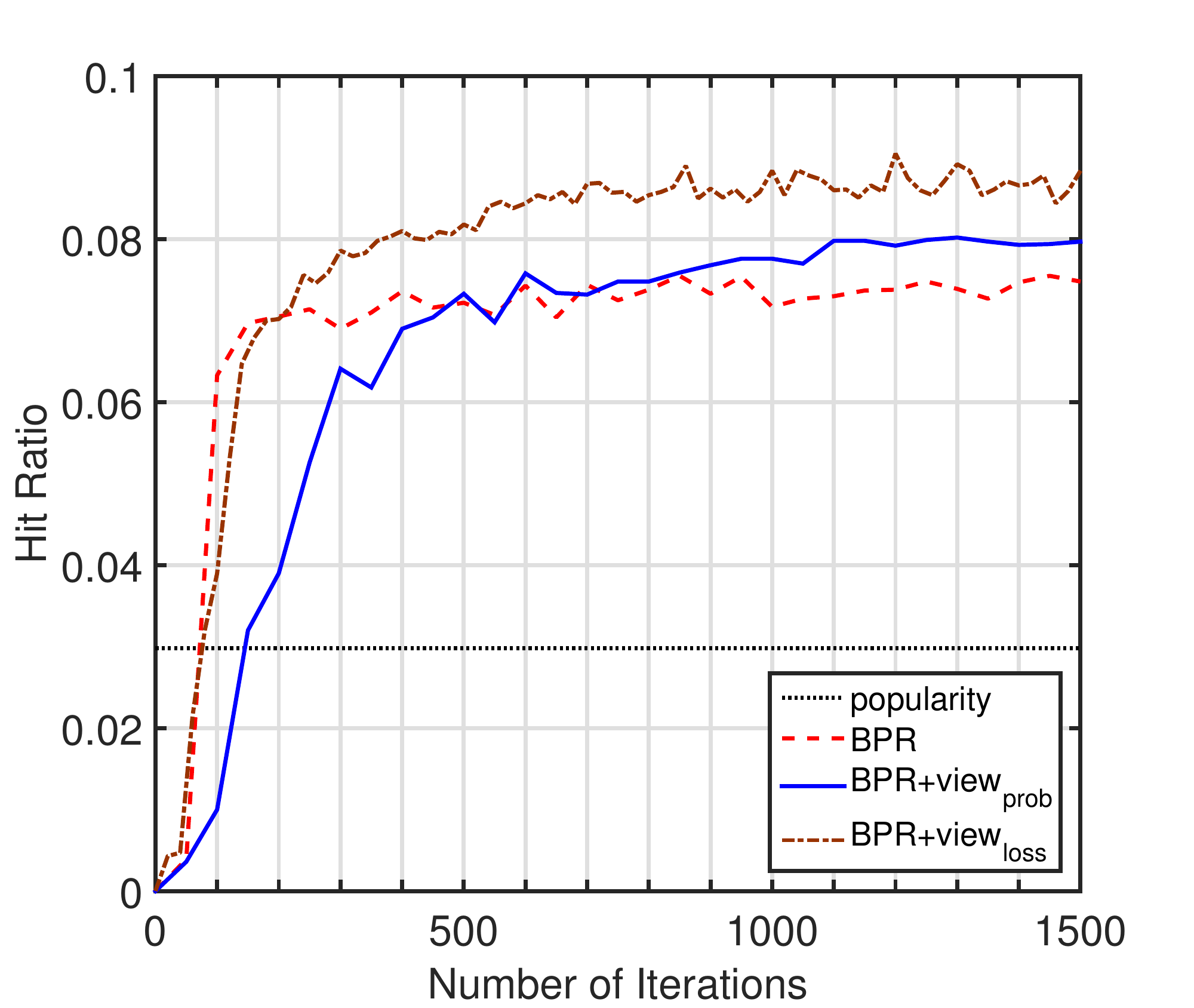}}
\hfill\hspace{-2em}
\subfloat[NDCG, Tmall-selected]{\includegraphics[width=.255\textwidth]{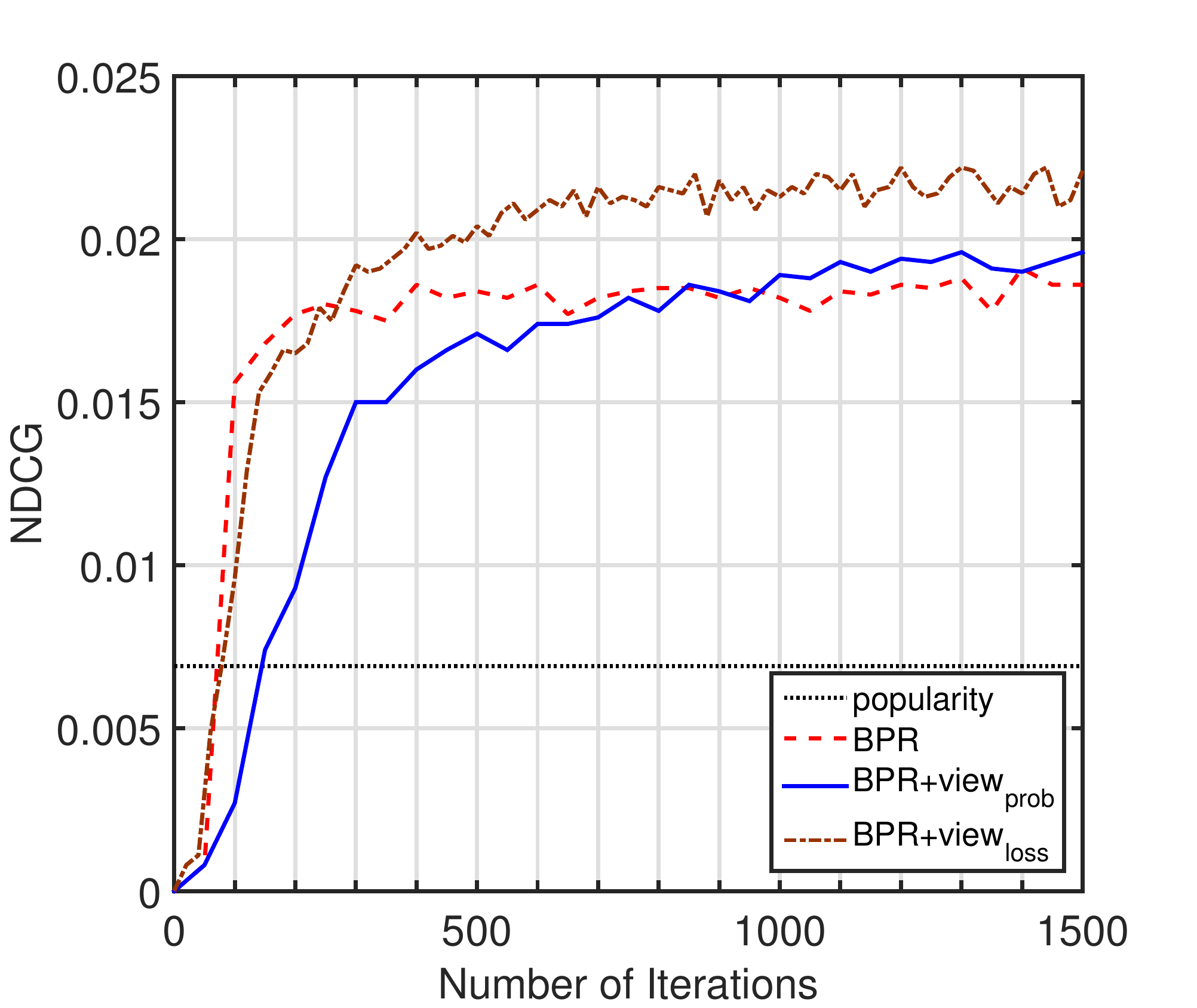}}
\caption{Performance comparison in each iteration~(\textit{BPR+view$_{loss}$}).}
\label{fig_perf_comp_2}
\end{figure}

\mypara{\textit{BPR+view$_{loss}$} vs. \textit{BPR+view$_{prob}$}.}
To evaluate our two proposed variants of BPR sampler, i.e., biased sampling scheme and weighted loss scheme, we look further into the comparison of \textit{BPR+view$_{prob}$} and \textit{BPR+view$_{loss}$} for every iteration, in Fig.~\ref{fig_perf_comp_2}. For Beibei, the relative improvement in terms of HR and NDCG are $1.29\%$ and $2.48\%$ respectively~($0.1436$ vs. $0.1422$ and $0.0327$ vs. $0.0321$, Table~\ref{tab_perf_comp}). Moreover, for Tmall-selected, we observe a relative improvement of $10.20\%$~($0.0884$ vs. $0.0807$) and $11.52\%$~($0.0221$ vs. $0.0199$) on two evaluation indexes, which indicates the stronger influence of viewing behavoirs on Tmall again. The obvious improvements demonstrates that considering three pairwise relations among the sampled item triple~(a purchased item, a viewed item and an unobserved item) can better describe both positive and negative signals of viewing behaviors. Even \textit{BPR+view$_{prob}$} outperforms vanilla \textit{BPR} and \textit{BPR-DNS}, it still has difficulty in treating view interactions as both positive and negative feedback in a single sampling.

\begin{figure}[t]
\centering
\subfloat[HR, Beibei]{\includegraphics[width=.255\textwidth]{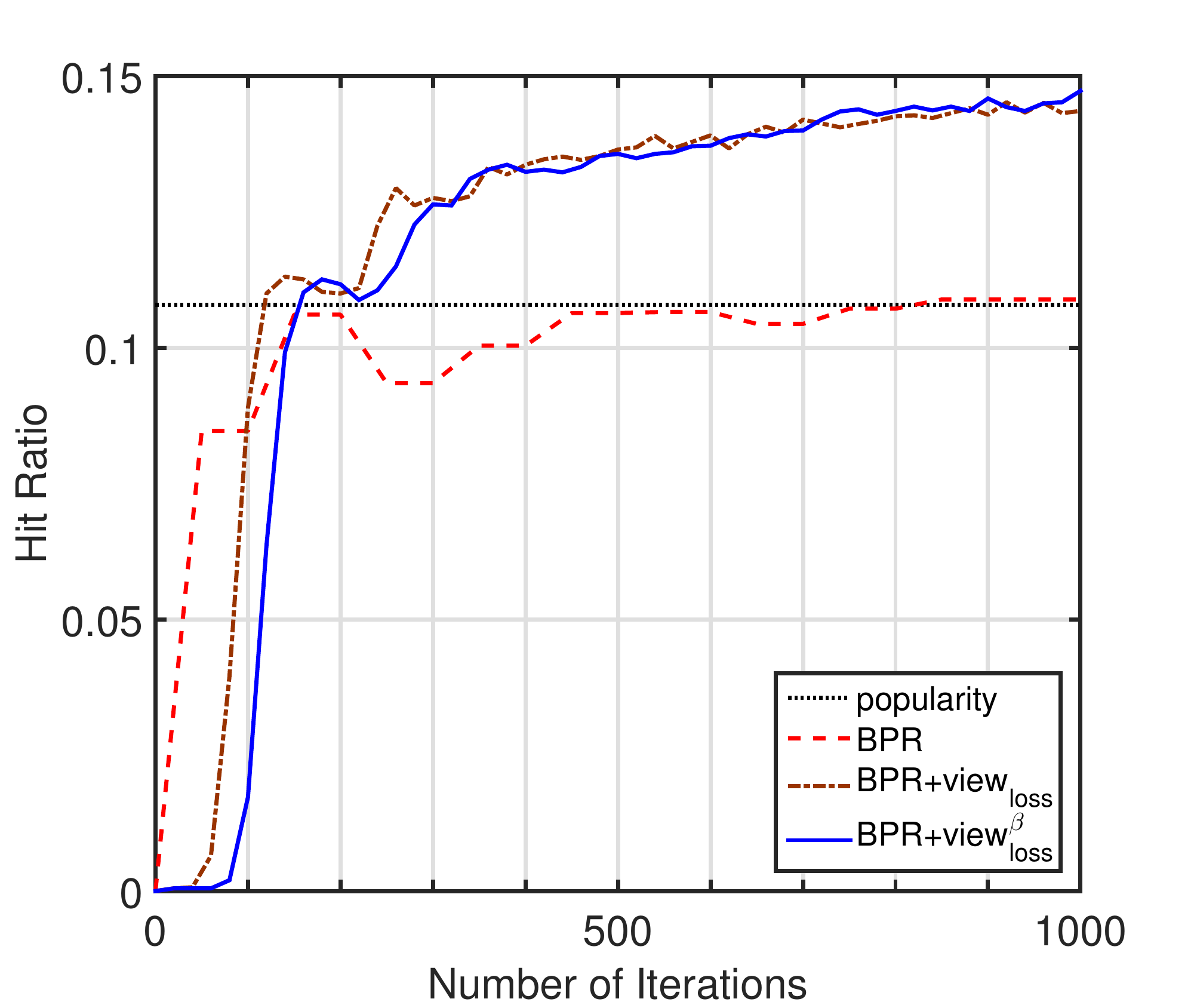}}
\hfill\hspace{-2em}
\subfloat[NDCG, Beibei]{\includegraphics[width=.255\textwidth]{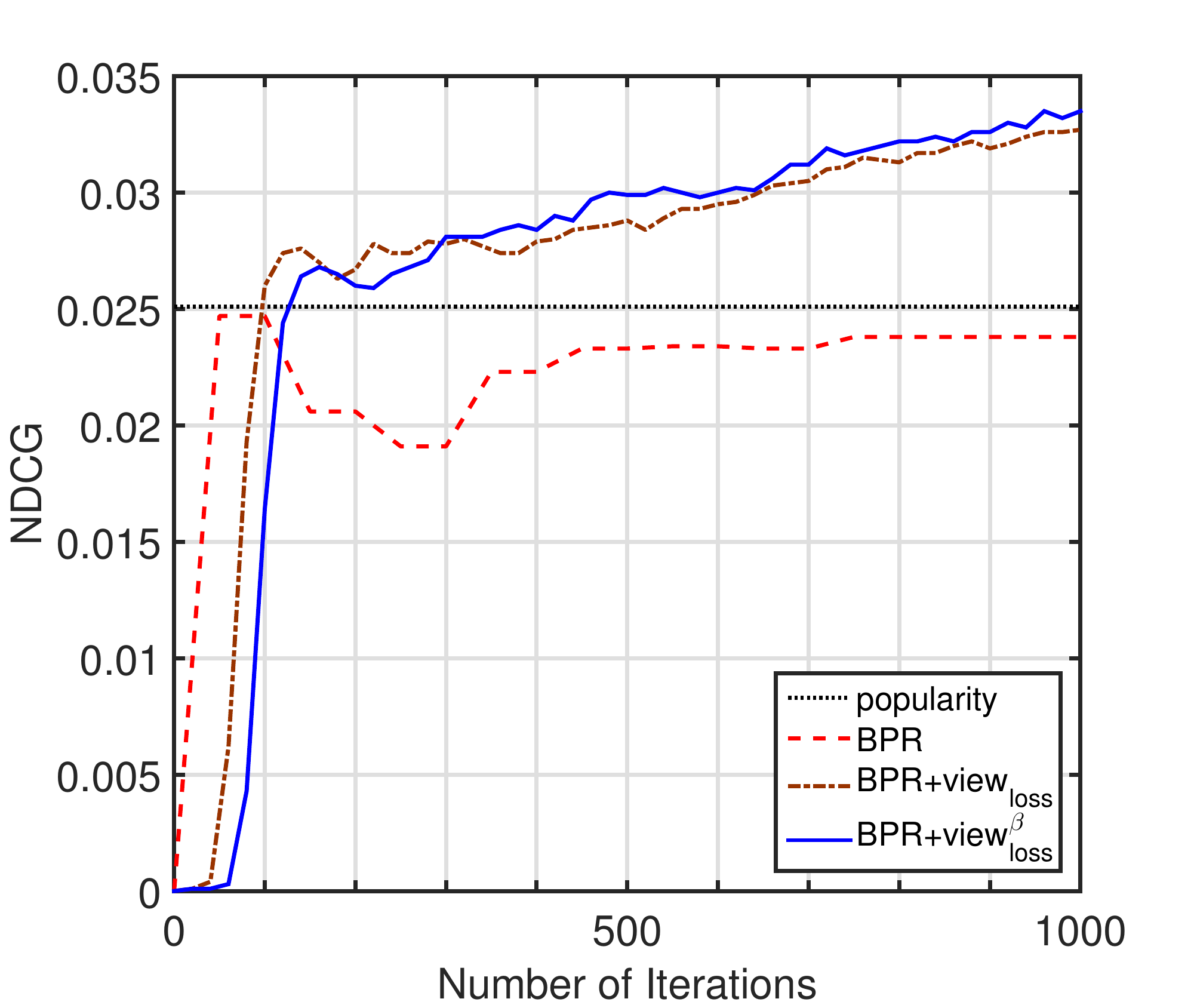}}
\caption{Performance comparison in each iteration~(\textit{BPR+view$^{\beta}_{loss}$}).}
\label{fig_perf_comp_3}
\end{figure}

\mypara{\textit{BPR+view$_{loss}$} vs. \textit{BPR+view$^{\beta}_{prob}$}.}
Finally, we compare the performance of \textit{BPR+view$_{loss}$} and \textit{BPR+view$^{\beta}_{loss}$} in Fig.~\ref{fig_perf_comp_3} to evaluate the efficacy of user-oriented weighting scheme. On Beibei dataset, by imposing personalized weighting strategy, \textit{BPR+view$^{\beta}_{loss}$} achieves a further relative improvement of $3.41\%$~($0.1473$ vs. $0.1436$)  and $3.31\%$~($0.0335$ vs. $0.0327$)  w.r.t. HR and NDCG, which proves our intuition that view interactions are stronger negative feedback for users with larger view-purchase ratio. 

To summarize, modelled as an intermediate feedback, users' viewed interactions can play an important role in learning a more precise user preference to improve recommendation performance. Compared with integrating viewing signal through a biased sampler, simultaneously learning two-fold semantics of viewing signal in each update step performs much better. By taking into account the effect of users' online-shopping habits, we design a user-oriented weighting scheme which achieves further improvements. 

\section{Conclusion and Future Work}
\label{sec:future}
This paper studied the problem of improving BPR sampler in implicit feedback recommender systems. First, we have demonstrated that sampling negative items from the whole space is unnecessary for BPR. Then, to further improve BPR sampler's ability of learning user preference, we propose an enhanced sampler that encodes two-fold semantics in user's viewing behaviors. With these design, our improved BPR sampler is able to achieve higher accuracy.

In the future, we will design an adaptive sampler to leverage view data and other implicit feedback more sufficiently. 
This work has focused on collaborative filtering setting, which only leverages the feedback data and is mostly used in the candidate selection stage of industrial recommender systems~\cite{wang2018path}. In future, we will focus more on the ranking stage, integrating view data into generic feature-based models, such as expressive neural factorization machines~\cite{NFM} and more explainable tree-enhanced embedding model~\cite{TEM}.

\newpage
\bibliographystyle{abbrv}
\bibliography{sigproc} 

% biography section
\begin{IEEEbiography}[{\includegraphics[width=1.0in,height=1.5in,clip,keepaspectratio]{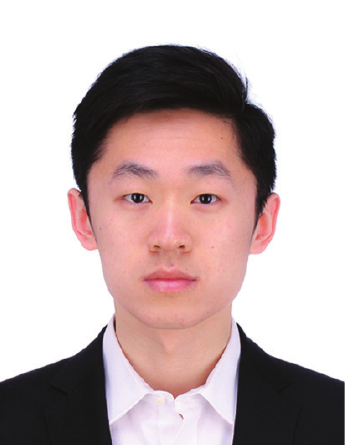}}]{Jingtao Ding} received the B.S. degrees in electronic engineering from Tsinghua University, Beijing, China, in 2015. He is currently pursuing the Ph.D. degree with the Department of Electronic Engineering, Tsinghua University.
His research interests include mobile computing, mobile data mining and user behavior modeling. \end{IEEEbiography}

\begin{IEEEbiography}
[{\includegraphics[width=1.0in,height=1.5in,clip,keepaspectratio]{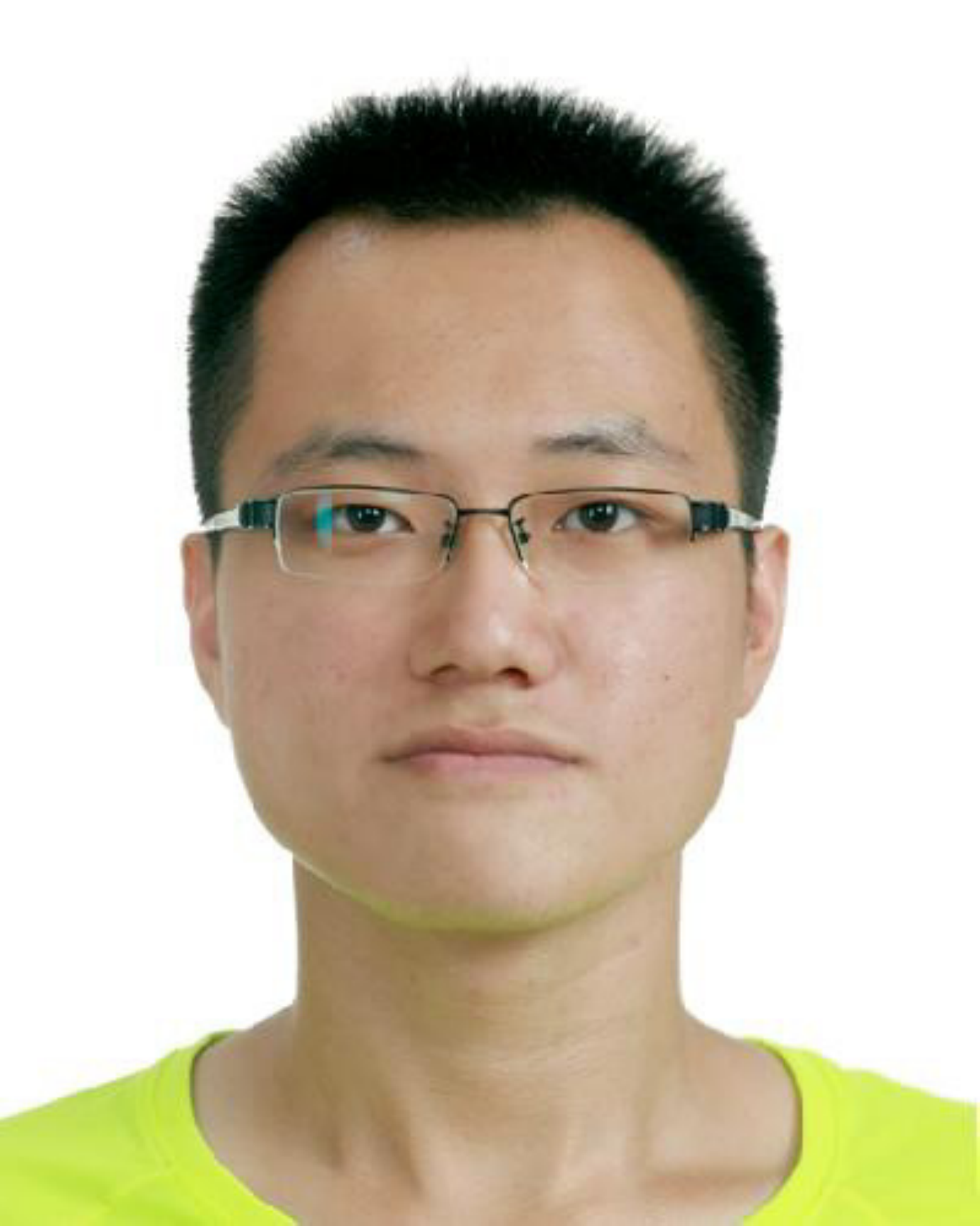}}]{Guanghui Yu} is an undergraduate student in electronic engineering department of Tsinghua University, Beijing, China. His research interests include user behavior modelling. \end{IEEEbiography}

\begin{IEEEbiography}
[{\includegraphics[width=1.0in,height=1.5in,clip,keepaspectratio]{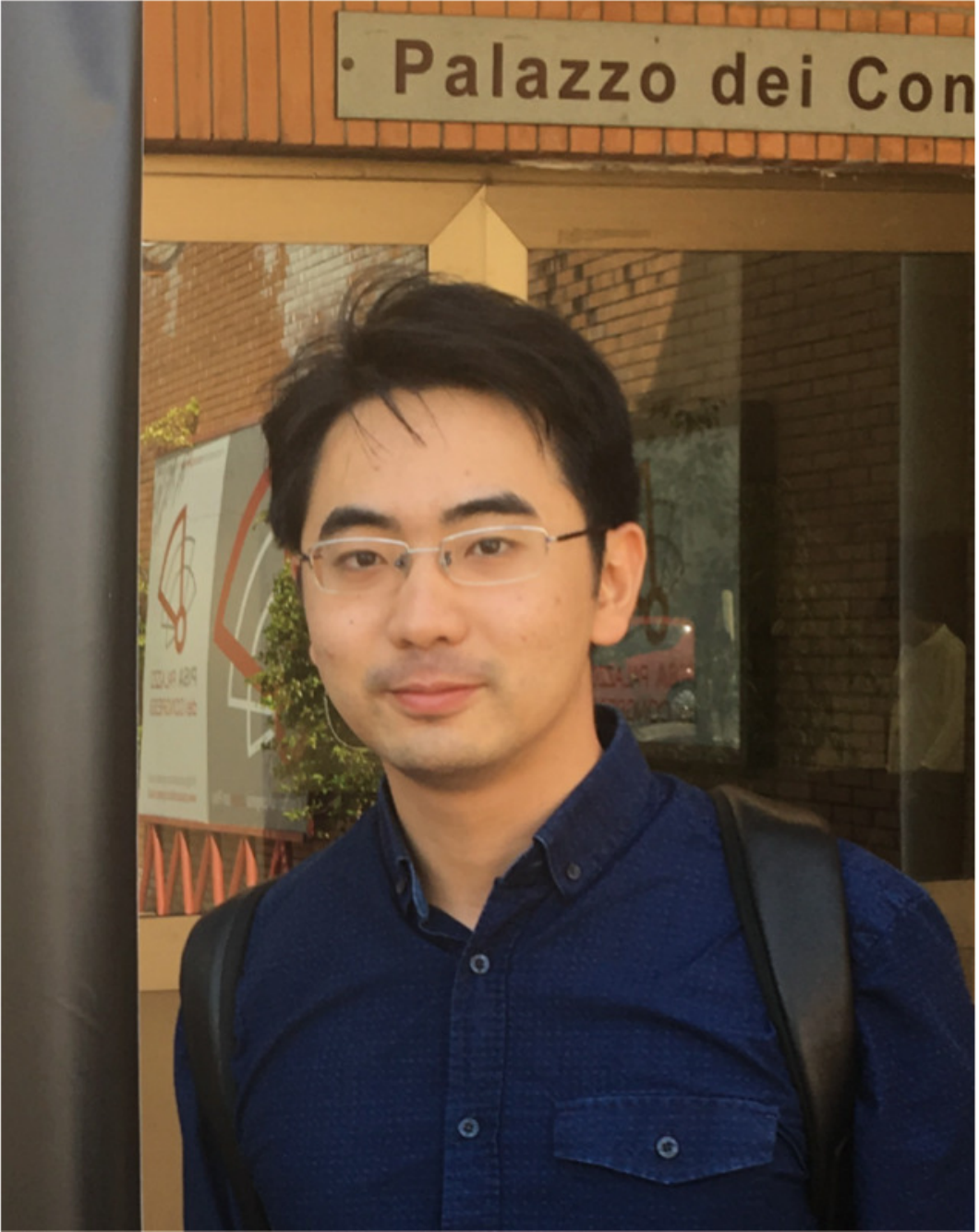}}]{Xiangnan He} is currently a research fellow with School of Computing, National University of Singapore (NUS). He received his Ph.D. in Computer Science from NUS. His research interests span recommender system, information retrieval, and multi-media processing. He has over 20 publications appeared in several top conferences
such as SIGIR, WWW, MM, CIKM, and IJCAI, and journals including TKDE, TOIS, and TMM. His work on recommender system has received the Best Paper Award Honorable Mention
of ACM SIGIR 2016. Moreover, he has served as the PC member for the prestigious conferences including SIGIR, WWW, MM, AAAI, IJCAI, WSDM, CIKM and EMNLP, and the regular reviewer for prestigious journals including TKDE, TOIS, TKDD, TMM etc. \end{IEEEbiography}

\begin{IEEEbiography}[{\includegraphics[width=1.0in,height=1.5in,clip,keepaspectratio]{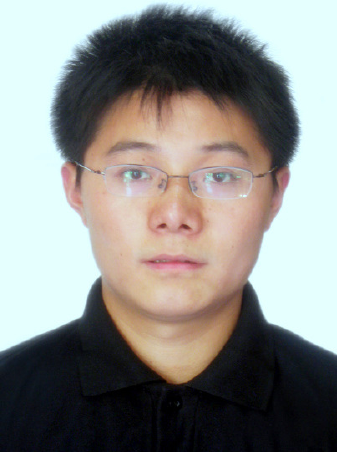}}]{Yong Li}
(M'2009-SM'2016) received the B.S. and Ph.D degree in Huazhong University of Science and Technology and Tsinghua University in 2007 and 2012, respectively. During 2012 and 2013, he was a Visiting Research Associate with Telekom Innovation Laboratories and Hong Kong University of Science and Technology respectively. During 2013 to 2014, he was a Visiting Scientist with the University of Miami. He is currently a Faculty Member of the Department of Electronic Engineering, Tsinghua University. His research interests are in the areas of Mobile Computing and Social Networks, Urban Computing and Vehicular Networks, and Network Science and Future Internet. 
Dr. Li has served as General Chair, Technical Program Committee (TPC) Chair, and TPC Member for several international workshops and conferences. He is currently the Associate Editor of Journal of Communications and Networking and EURASIP Journal of Wireless Communications and Networking \end{IEEEbiography}

\begin{IEEEbiography}[{\includegraphics[width=1.0in,height=1.5in,clip,keepaspectratio]{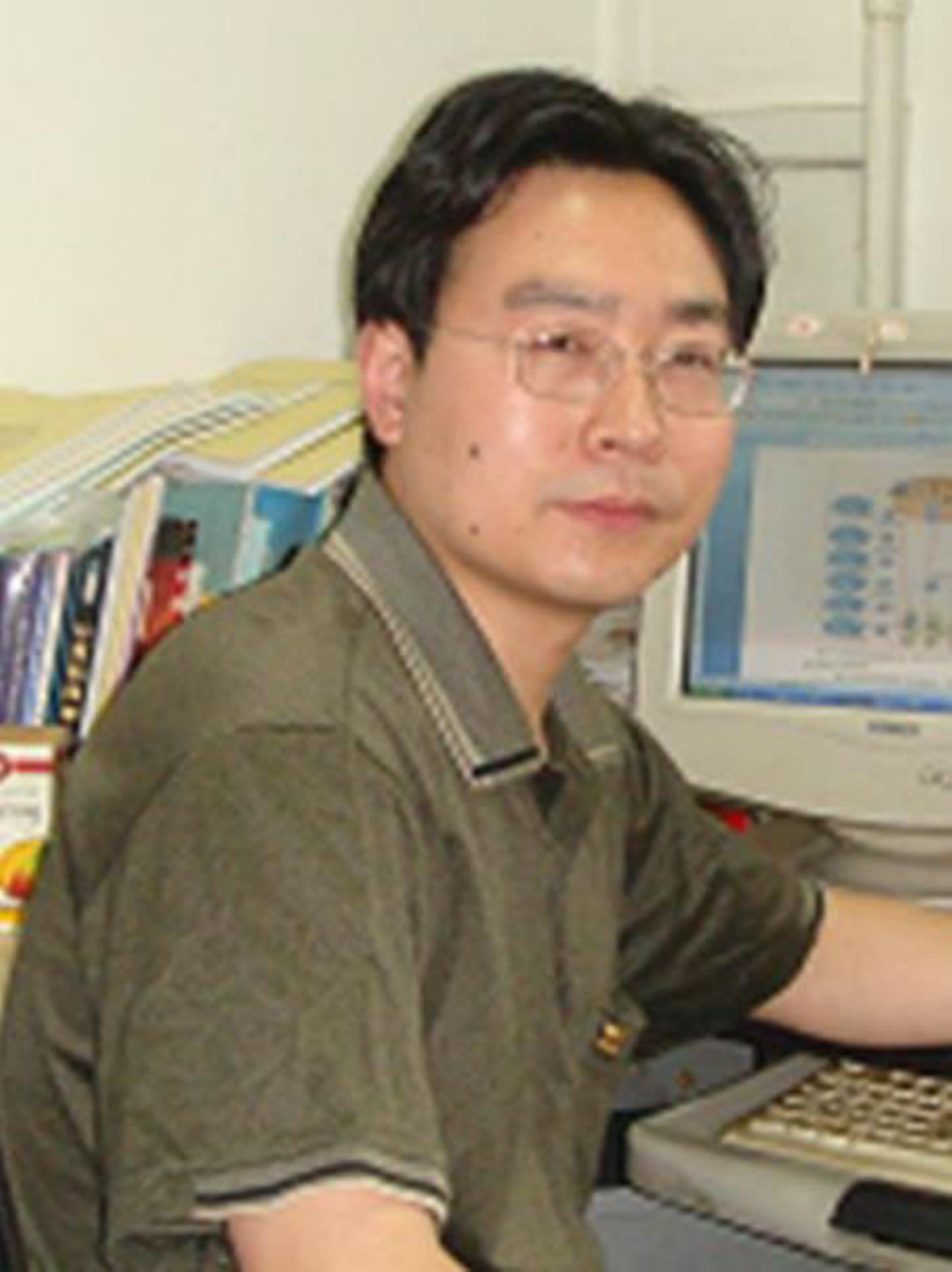}}]{Depeng Jin}
 (M'2009) received his B.S. and Ph.D. degrees from Tsinghua University,
 Beijing, China, in 1995 and 1999 respectively both in electronics
 engineering. Now he is an associate professor at Tsinghua University
 and vice chair of Department of Electronic Engineering. Dr. Jin was
 awarded National Scientific and Technological Innovation Prize
 (Second Class) in 2002. His research fields include
 telecommunications, high-speed networks, ASIC design and future
 internet architecture.
\end{IEEEbiography}

\end{document}